\documentclass[sn-mathphys]{sn-jnl}

\jyear{2021}%

\theoremstyle{thmstyleone}%
\theoremstyle{thmstyletwo}%

\theoremstyle{thmstylethree}%

\begin{document}

\title[Article Title]{Photon emission in the graphene under the action of a quasiconstant
external electric field}


\author*[1,2]{\fnm{S.P.} \sur{Gavrilov}}\email{gavrilovsergeyp@yahoo.com, gavrilovsp@herzen.spb.ru}

\author[1,3,4]{\fnm{D.M.} \sur{Gitman}}\email{gitman@if.usp.br}
\equalcont{These authors contributed equally to this work.}

\affil*[1]{\orgdiv{Department  of Physics}, \orgname{Tomsk State University},  \orgaddress{\city{Tomsk}, \postcode{634050}, \country{Russia}}}

\affil[2]{\orgdiv{Department  of General and Experimental Physics}, \orgname{Herzen State
Pedagogical University of Russia}, \orgaddress{\street{48 Moyka embankment}, \city{St. Petersburg}, \postcode{191186},  \country{Russia}}}

\affil[3]{\orgname{P.N. Lebedev Physical Institute}, \orgaddress{\street{53 Leninskiy prospect}, \city{Moscow}, \postcode{119991},  \country{Russia}}}

\affil[4]{\orgdiv{Institute of Physics}, \orgname{University of S\~{a}o Paulo}, \orgaddress{\street{CP 66318}, \city{S\~{a}o Paulo}, \postcode{05315-970}, \state{SP}, \country{Brazil}}}


\abstract{Following a nonperturbative formulation of strong-field QED developed in our
earlier works, and using the Dirac model of the graphene, we construct a
reduced QED$_{3,2}$\ to describe one species of the Dirac fermions in the
graphene interacting with an external electric field and photons. On this
base, we consider the photon emission in this model and construct closed
formulas for the total probabilities. Using the derived formulas, we study
probabilities for the photon emission by an electron and for the photon
emission accompanying the vacuum instability in the quasiconstant electric
field that acts in the graphene plane during the time interval $T$.\ We
study angular and polarization distribution of the emission as well as
emission characteristics in a high frequency and low frequency
approximations. We analyze the applicability of the presented calculations
to the graphene physics in laboratory conditions. In fact, we are talking
about a possible observation of the Schwinger effect in these conditions.}




\maketitle

\section{Introduction}

\label{sec1}

Graphene and similar nanostructures (topological insulators, etc.) belong to
the class of so-called Dirac semimetals, the theoretical and experimental
study of which has recently received much attention. In particular, this is
due to the hopes for possible applications of these structures in the next
generation of semiconducting devices. It should be noted that already in the
early theoretical works \cite{semenoff,Geim2007,NKats2007} it was indicated
that the physics of the graphene is not only described under certain
conditions by QFT (quantum field theory), but can be a fertile area for the
latter where its validity within extreme limits can be verified in
laboratory conditions. This possibility is related to the fact that
low-energy single-electron motion in graphene monolayers (at the charge
neutrality point) and similar nanostructures is described by the Dirac
model, being a $2+1$ dimensional version of massless QED (quantum
electrodynamics) with Fermi velocity $v_{F}\simeq 10^{6}\mathrm{m/s}$
playing the role of the light speed in the relativistic dynamics of the
corresponding Dirac fermions \cite{semenoff}, see also the review \cite%
{gusynin} for more details. Such a model is usually called reduced QED$%
_{3,2} $. It should be noted that in the QED$_{3,2}$ model the
electromagnetic field itself is not confined to the graphene plane $z=0$,
but rather propagates (with the speed of light $c$) according to
corresponding classical or quantum equations in the ambient $3+1$\
dimensional space-time. The electromagnetic field couples minimally to
electrons situated on the graphene plane. Thus, we note once again, that in
the QED$_{3,2}$ there are two distinct velocities, one of charged particles
(Dirac fermions) and another one of the electromagnetic fields. Since the
Dirac fermions in the model are considered as almost massless, any
low-frequency electric field is for them supercritical (the so-called
Schwinger critical field $E_{\mathrm{c}}=m^{2}c^{3}/e\hslash $ is almost
zero). The latter fact allows one a laboratory verifying QED predictions for
superstrong fields, in particular, real studying the Schwinger effect. From
the theoretical point of view, what has been said means that the vacuum
state in the model is sometimes unstable with respect to the Dirac fermion
creation, such that the interaction with electric-like external field must
be taken into account nonperturbatively. That is why the standard theory of
the photon emission represented in QED text-books cannot help. From our
point of view, adequate nonperturbative calculations with respect to the
external field can be done using a general approach to QED with strong
external fields \cite{GenTheory1,GenTheory2,GenTheory3,FGS91} (based on the
existence of special exact solutions of the Dirac equation with these
fields) and its further development \cite{277,357}.\textrm{\ }Thus,\textrm{\
}the QED$_{3,2}$ model with a part of electric-like electromagnetic field
considering as an external classical one must always be treated by the above
mentioned nonperturbative methods. Note that the effects due to
high-frequency electromagnetic fields, which are often considered in
connection with the optical response of Dirac fermions in graphene, do not
require the use of the mentioned nonperturbative methods. In the QED$_{3,2}$
model, there are actually two species of fermions corresponding to
excitations about two distinct Dirac points in the Brillouin zone of the
graphene. Taking into account the presence of two spin polarization of
excitations of each kind, we have, in fact, four species of fermions in the
model. Calculations of mean values can be done for one of the specie with
some further extension to four species using the degeneracy factor $N_{f}=4$.

Until now, in $3+1$ QED, there\textrm{\ }were presented various
nonperturbative calculations of zero order processes in the framework of the
general approach \cite{GenTheory1,GenTheory2,GenTheory3,FGS91,277,357}; for
example, see Refs. \cite{GG96,AdoGavGit17} and references therein. These are
processes of charged particle scattering, and processes of charge particle
creation and annihilation related to the vacuum instability. In QED$_{3,2}$,
the graphene conductivity modification due to the particle creation by
external constant electric field (the Schwinger effect) was calculated as a
zero order process in Ref. \cite{GavGitY12}. Processes involving photon
emission and annihilation in the presence of the vacuum instability are
processes of higher order in radiative corrections. Their study is
technically more complicated then the study of zero order processes.
Nevertheless, recently, there appeared publications devoted to the photon
emission in the graphene due to external electric field in the framework of
the Dirac model. In particular, a free electron-hole recombination was
studied in Refs.\cite{mecklenburg} and \cite{lewkowicz-11} for the thermal
equilibrium. A discussion of the photon emission by charged carries in the
graphene due to constant uniform electric field was published in Ref. \cite%
{Yok14}. Due to the recent detection of an optical radiation{\large \ }in
the graphene\emph{\Huge \ }accompanying the creation of electron-hole pairs
by a terahertz pulse \cite{emis-exp17,emis-exp21}, it becomes possible to
make a comparison of the corresponding theoretical calculations with
experiments. It must be said that the emission of a photon by an electron
moving in a constant electric field in $3+1$ dimensions was studied first
nonperturbatively by Nikishov \cite{nikishov,nikishov79}.

Peculiarities of physics of the graphene allows one studying the Schwinger
effect in laboratory conditions. Theoretical calculations presented in the
work \cite{GavGitY12}{\large \ }and their comparison with experimentally
observed results of the $dc$ conductivity in the graphene near the Dirac
point testify in favor of the fact that it is the Schwinger effect that
determines the nature of the conductivity. In addition, it has been found
that the radiation of a time-dependent mean current, forming the
backreaction to the electric field on the graphene plane,{\large \ }is
emitted to the three-dimensional space in the form of linearly polarized and
of very low frequency plane electromagnetic waves.{\large \ }However, an
observation of such waves and the $dc$ conductivity is not a simple task in
the presence of the background noise in the vicinity of the graphene sample.%
{\large \ }We believe that the emission and absorption of high-frequency
photons accompanied the electronic quantum transport in the graphene are
more realistic for possible experimental observations.

We stress that general equations allowing nonperturbative calculations of
the higher order processes in strong-field QED are clearly formulated in the
Refs. \cite{GenTheory1,GenTheory2,GenTheory3,FGS91,277,357}. In the present
study, we specify these equations for the above described QED$_{3,2}$ model
and with their help we consider{\large \ }processes of photon emission by
the Dirac excitations in the graphene subjected by external constant
electric field. In these calculations effects of the vacuum instability are
taken into account exactly, such that we study the process of the photon
emission which is accompanied by creation from the vacuum additional Dirac
excitations.

In contrast to the works known to us, in this article we consider effects in
the QED$_{3,2}$ model with an intense external electric field, which is
uniform and slowly varying, and which we interpret as a macroscopic external
field. These effect differ principally from effects arising in magnetic-like
fields or in fields of high-frequency electromagnetic waves. In condensed
matter the Dirac model is used primarily in the context of the relativistic
quantum mechanics, or in the framework of the Matsubara's imaginary time
formalism of QFT, where electrons are assumed to be in thermal equilibrium;
see, e.g. Refs. \cite%
{castroneto1,castroneto2,castroneto3,castroneto4,castroneto5,dassarma,Kotov+etal12,three-loop-QED,magn11,MirS15,kats20}%
. However, a macroscopic electric field acting on charged particles may
destroy their thermal equilibrium, such that conclusions based on the latter
assumption may be not correct.

The article is organized as follows: In Sect. \ref{S2} we, following the
general theory \cite{GenTheory1,GenTheory2,GenTheory3,FGS91}, construct the
reduced QED$_{3,2}$\ to describe one species of the Dirac fermions in the
graphene interacting with an external electric field and photons. The
required basic elements that we need to describe zero-order processes with
respect to the electron-photon interaction are derived from Ref.~\cite%
{GavGitY12}. Then, we consider the photon emission in this model and
construct closed formulas for the total probabilities. In order to find the
corresponding mean values in real graphene, results obtained for one species
of the Dirac fermions must be multiplied by the factor $N_{f}=4$ (the number
of all charged species). In Sect. \ref{S3}, we apply the developed
formulation to calculating probabilities of the one-photon emission by an
electron and of the one-photon emission accompanying the vacuum instability
in a quasiconstant electric field that acts in the graphene plane during the
time interval $T$.\ In Sect. \ref{S4}, we analyze the obtained emissions
characteristics in a high frequency approximation. We study angular and
polarization distribution of the emissions. The low frequency approximation
is considered in Appendix \ref{App2}. We analyze conditions of the
applicability of the presented calculations in possible experimental
conditions. In the last Sect. \ref{S5}, we summarize the main results of the
present work. Some useful mathematical details are placed in Appendix \ref%
{App1}.

\section{The photon emission in the graphene in the framework of the QED$%
_{3,2}$ model}

\label{S2}

\subsection{General}

\label{SS2.1}

In this section we consider general equations that will be used by us
further to study the photon emission in a flat graphene monolayer in the
framework of the nonperturbative approach \cite%
{GenTheory1,GenTheory2,GenTheory3,FGS91} applied to the QED$_{3,2}$ model
described schematically above.

We consider an infinite flat graphene sample to which an uniform electric
field is applied, directed constantly along the axis $x$\ on the plane of
the sample. We assume that the applied field is a strong external
macroscopic low-frequency electric-like field that can treated as a
quasiconstant one. We consider the case of zero temperature and chemical
potential (i.e., at the charge neutrality point), so that the Dirac model
can be used near the Dirac point.

As was already said the graphene sample in subjected to the action of a
strong external macroscopic low-frequency electric-like field, some
suppositions about which were already mentioned above. This field is
parallel to the graphene plane, $z=0$. By $\mathbf{r}=\left( x,y\right) $ we
denote the two-dimensional position vector on the graphene plane. In which
follows we use boldface symbols for any two-dimensional vectors in $z=0$
plane. The electromagnetic field couples minimally to the current of the
Dirac fermions in the graphene plane. The external field can be given by
two-dimensional vector potential $\mathbf{A}^{\mathrm{ext}}(t,\mathbf{r,}%
z)=\left( A_{x}^{\mathrm{ext}},A_{y}^{\mathrm{ext}}\right) $ (the scalar
potential is chosen to be zero, $A_{0}=0)$. In the model under
consideration, charged particles of each kind in the graphene are described
by the Dirac field which is two component spinor $\psi _{\alpha }(t,\mathbf{r%
})$, $\alpha =1,2$ on $2+1$ dimension. In this dimension, the algebra of the
corresponding $\gamma $-matrices has two inequivalent representations,
\begin{equation}
\gamma ^{0}=\sigma ^{3}\,,\;\gamma ^{1}=i\sigma ^{2}\,,\;\gamma
^{2}=-i\varsigma \sigma ^{1}\,,  \label{gamma}
\end{equation}%
where the $\sigma ^{j}$ are Pauli matrices, and by $\varsigma =\pm 1$
inequivalent representations are labeled. Distinct (pseudo spin)
representations are associated with each Dirac point. For all integral
quantities, since intervalley scattering can be neglected, the presence of
two valleys related to each $\varsigma =\pm 1$\ inequivalent representation
is taken into account simply by multiplying by introducing the degeneracy
factor $2$.{\large \ }Taking into account the spin degeneracy factor $2$,
the total number of different species of Dirac fermions is $N_{f}=4$.{\large %
\ }In order to find mean values of a physical quantity in the graphene, a
mean value obtained for one species are multiplied by{\large \ }$N_{f}\ $.
{\large \ }Remembering the origin of the Dirac model for the graphene
description (see \cite{semenoff}), we believe that each component $\psi
_{\alpha }(t,\mathbf{r})$ of the Dirac spinor is a projection of{\Large \ }a
Schr\"{o}dinger wave function $\phi _{\alpha }(t,\mathbf{r},z)$ in $3+1$
dimensions with a support in a specific sublattice of the honeycomb lattice
of the graphene. These wave functions can be represented as:
\begin{equation}
\phi _{\alpha }(t,\mathbf{r},z)=\psi _{\alpha }(t,\mathbf{r})\varphi
(z)\,e^{ip_{z}z/\hbar }\,,  \label{eq:2d-3d-map}
\end{equation}%
where the function $\varphi (z)$ describes the width of the graphene. A
detailed description of $\varphi (z)$ is not necessary for our purposes,
except for the fact that it decays rapidly outside the $xy$ plane and is
normalized according to $\int dz {\|\varphi (z) \|}^{2}$. In which follows
we assume the usual dipole approximation so that the exponential in Eq. (\ref%
{eq:2d-3d-map}) is approximated by the zeroth-order constant term. In this
approximation we replace{\Large \ }$A^{\mathrm{ext}}(t,\mathbf{r},z)$\ by
its value $\mathbf{A}^{\mathrm{ext}}(t,\mathbf{r,}0)$ at $z=0.$\ Then we can
simplify the notation as follows:{\Large \ }$\mathbf{A}^{\mathrm{ext}}(t,%
\mathbf{r,}0)=\mathbf{A}^{\mathrm{ext}}(t,\mathbf{r})$.{\Large \ }We allow
the graphene sheet to have a global momentum $p_{z}$ along the $z$ axis, in
order to account for the possibility of a momentum transfer in this
direction with respect to some external system. The Dirac equation with an
external field that couples minimally to electrons on graphene plane reads:%
\begin{eqnarray}
&&i\hbar \,\partial _{t}\psi (t,\mathbf{r})=H^{\mathrm{ext}}\psi (t,\mathbf{r%
})\,,  \notag \\
&&H^{\mathrm{ext}}=v_{F}\gamma ^{0}\left\{ \boldsymbol{\gamma }\left[
\mathbf{p}+\frac{e}{c}\mathbf{A}^{\mathrm{ext}}(t,\mathbf{r})\right]
+mv_{F}\right\} \ ,  \label{Dext}
\end{eqnarray}%
where $\mathbf{p=}\left( p_{x},p_{y}\right) $ is the in-plane component of
the momentum operator, $\boldsymbol{\gamma =\ }\left( \gamma ^{1},\gamma
^{2}\right) $, $\gamma $-matrices satisfy the standard anticommutation
relations $\left[ \gamma ^{\mu },\gamma ^{\nu }\right] _{+}=2\eta ^{\mu \nu
} $, $\eta _{\mu \nu }=\text{diag}\,(+1,-1,-1)$\emph{, }$\mu ,\nu =0,1,2$, \
and $e>0$ is the absolute value of the electron charge.

In Eq. (\ref{Dext}) a mass term in the Hamiltonian $H^{\mathrm{ext}}$ is
introduced for one to be able to generalize the consideration to the
presence of the possible mass gap $\Delta \varepsilon =mv_{F}^{2}$. Such a
mass gap in the graphene band structure can appear in different ways. One of
the examples is given by graphene nanoribbons, see \cite{dassarma} for a
review. However, in our consideration below, we set $m=0$.

Dirac Heisenberg operators $\hat{\Psi}\left( t,\mathbf{r}\right) $ and $\hat{%
\Psi}^{\dag }\left( t,\mathbf{r}\right) $ are assigned to the Dirac fields $%
\psi (t,\mathbf{r})$ and $\psi ^{\dag }(t,\mathbf{r})$. These fields obey
both the Dirac equation (\ref{Dext}) with the potential $\mathbf{A}^{\mathrm{%
ext}}(t,\mathbf{r})$] and the following nonvanishing equal time
anticommutation relations:%
\begin{equation}
\left[ \hat{\Psi}\left( t,\mathbf{r}\right) ,\hat{\Psi}\left( t,\mathbf{r}%
^{\prime }\right) \right] _{+}\ ,\ \left[ \hat{\Psi}\left( t,\mathbf{r}%
\right) ,\hat{\Psi}^{\dag }\left( t,\mathbf{r}^{\prime }\right) \right]
_{+}=\hbar \delta ^{\left( 2\right) }\left( \mathbf{r}-\mathbf{r}^{\prime
}\right) .  \label{f4}
\end{equation}%
The quantized free electromagnetic field is described by{\Large \ }%
two-dimensional operators of vector potential{\Large \ }$\mathbf{\hat{A}}(t,%
\mathbf{r,}z)$. As for the classical potentials, the dipole approximation
allows us to replace $\mathbf{\hat{A}}(t,\mathbf{r,}z)$ by its value $%
\mathbf{\hat{A}}(t,\mathbf{r})=\mathbf{\hat{A}}(t,\mathbf{r,}0)$.

The total quantum Hamiltonian of the model reads:%
\begin{eqnarray}
&&\widehat{\mathcal{H}}\left( t\right) =\widehat{\mathcal{H}}_{\mathrm{e,}%
\mathbf{A}^{\mathrm{ext}}}+\widehat{\mathcal{H}}_{\mathrm{e,\gamma }}\ +%
\widehat{\mathcal{H}}_{\mathrm{\gamma }}\ ,  \notag \\
&&\widehat{\mathcal{H}}_{\mathrm{e,}\mathbf{A}^{\mathrm{ext}}}=\int \hat{\Psi%
}^{\dag }\left( t,\mathbf{r}\right) H^{\mathrm{ext}}\hat{\Psi}\left( t,%
\mathbf{r}\right) d\mathbf{r},  \notag \\
&&\widehat{\mathcal{H}}_{\mathrm{e,\gamma }}=-\int \mathbf{\widehat{\mathbf{j%
}}(}t\mathbf{,\mathbf{r})\hat{A}}(t,\mathbf{r})d\mathbf{r}\ ,\;\mathbf{%
\widehat{\mathbf{j}}(}t\mathbf{,\mathbf{r})}=-\frac{ev_{F}}{2c}\left[ \hat{%
\Psi}^{\dag }\left( t,\mathbf{\mathbf{r}}\right) ,\gamma ^{0}\mathbf{\gamma }%
\hat{\Psi}\left( t,\mathbf{\mathbf{r}}\right) \right] _{-}\;,  \label{f3}
\end{eqnarray}%
where$\ \widehat{\mathcal{H}}_{\mathrm{e,}\mathbf{A}^{\mathrm{ext}}}$ is the
Hamiltonian of charged particles interacting with an external electric-like
field given by the time-dependent potential $\mathbf{A}^{\mathrm{ext}}(t,%
\mathbf{r})$, $\widehat{\mathcal{H}}_{\mathrm{e,\gamma }}$ is the
Hamiltonian of the electron-photon interaction, and $\widehat{\mathcal{H}}_{%
\mathrm{\gamma }}$ is the free photon Hamiltonian. The integral on the
graphene plane is taken over an area{\Large \ }$S${\Large . }We assume that
the area $S$\ is sufficiently large to be macroscopic then boundary effects
can be neglected.

The decomposition of the operator $\mathbf{\hat{A}}(t,\mathbf{r})$ in terms
of annihilation and creation operators of free photons, $C_{\mathbf{K}%
\vartheta }$ and $C_{\mathbf{K}\vartheta }^{\dagger }$ reads:
\begin{equation}
\mathbf{\hat{A}}(t,\mathbf{r})=c\sum_{\mathbf{K,}\vartheta }\sqrt{\frac{2\pi
\hbar }{\varepsilon V\omega }}\boldsymbol{\epsilon }_{\mathbf{K}\vartheta }%
\left[ C_{\mathbf{K}\vartheta }\,e^{i(\mathbf{kr}-\omega t)}+C_{\mathbf{K}%
\vartheta }^{\dagger }\,e^{-i(\mathbf{kr}-\omega t)}\right] \,,
\label{eq:vector-potential}
\end{equation}%
where $\vartheta =1,2$ denotes a polarization index, $\boldsymbol{\epsilon }%
_{\mathbf{K}\vartheta }$ are mutual orthogonal unit polarization vectors
transversal to three-dimensional wave vector $\mathbf{K}=\left( \mathbf{k}%
,k_{z}\right) $. The two-dimensional vector $\mathbf{k=\ }\left(
k_{x},k_{y}\right) $ is a projection of $\mathbf{K}$ on the graphene plane, $%
\omega =cK,$ $K=\left\vert \mathbf{K}\right\vert $, $V$ is the volume of the
box regularization, and $\varepsilon $ is the relative permittivity (for the
graphene suspended in the vacuum $\varepsilon =1$).

\subsection{In- and out-states of charged particles\label{SS2.2}}

Following the general nonperturbative approach \cite%
{GenTheory1,GenTheory2,GenTheory3,FGS91} we have to construct the
corresponding \textrm{in}- and \textrm{out}-states. of charged particles of
all the kinds with the help of exact solutions of equation (\ref{Dext}) the
electric-like external field. As was already mentioned above, the external
field in the model is a slowly varying uniform electric-like field directed
along the axis $x$. It is assumed that for $t<t_{1}$ and for $t>t_{2}$, the
electric field is absent, therefore initial $\left\vert 0,\mathrm{in}%
\right\rangle _{\mathrm{e}}$ and final $\left\vert 0,\mathrm{out}%
\right\rangle _{\mathrm{e}}$ are vacuum state of free $\mathrm{in}$- and $%
\mathrm{out}$- charged particles, respectively. These vacua are different
due to a difference of initial and final values of external electromagnetic
field potentials. During the time interval $t_{2}$ $-t_{1}$ $=T$, the Dirac
field interacts with the external field. There exists a set of creation and
annihilation operators $a_{n}^{\dagger }(\mathrm{in})$ and $a_{n}(\mathrm{in}%
)$ of $\mathrm{in}$-particles (electrons), and operators $b_{n}^{\dagger }(%
\mathrm{in})$ and $b_{n}(\mathrm{in})$ of $\mathrm{in}$-antiparticles
(holes), at the same time there exists a set of creation and annihilation
operators $a_{n}^{\dagger }(\mathrm{out})$ and $a_{n}(\mathrm{out})$ of $%
\mathrm{out}$-electrons and operators $b_{n}^{\dagger }(\mathrm{out})$ and $%
b_{n}(\mathrm{out})$ of $\mathrm{out}$-holes,

\begin{eqnarray}
a_{n}\left( \mathrm{in}\right) \left\vert 0,\mathrm{in}\right\rangle _{%
\mathrm{e}} &=&b_{n}\left( \mathrm{in}\right) \left\vert 0,\mathrm{in}%
\right\rangle _{\mathrm{e}}=0,\;\forall n,  \notag \\
a_{n}\left( \mathrm{out}\right) \left\vert 0,\mathrm{out}\right\rangle _{%
\mathrm{e}} &=&b_{n}\left( \mathrm{out}\right) \left\vert 0,\mathrm{out}%
\right\rangle _{\mathrm{e}}=0,\;\forall n,  \label{f5}
\end{eqnarray}%
In both cases, by $n$ we denote complete sets of quantum numbers describing $%
\mathrm{in}$- and $\mathrm{out}$- charged particles. As will be seen further
in the case under consideration $n=\mathbf{p.}$ The $\mathrm{in}$- and $%
\mathrm{out}$-operators obey the nonzero anticommutation relations:
\begin{align*}
& [a_{n}(\mathrm{in}),a_{n^{\prime }}^{\dagger }(\mathrm{in})]_{+}=[a_{n}(%
\mathrm{out}),a_{n^{\prime }}^{\dagger }(\mathrm{out})]_{+}=\hbar \delta
_{n,n^{\prime }}\,, \\
& [b_{n}(\mathrm{in}),b_{n^{\prime }}^{\dagger }(\mathrm{in})]_{+}=[b_{n}(%
\mathrm{out}),b_{n^{\prime }}^{\dagger }(\mathrm{out})]_{+}=\hbar \delta
_{n,n^{\prime }}\,.
\end{align*}

The $\mathrm{in}$-operators are associated with a complete orthonormal set
of solutions $\left\{ \ _{\zeta }\psi _{n}\left( t,\mathbf{r}\right)
\right\} $ ($\zeta =+$ for electrons and $\zeta =-$ for holes) of the Dirac
equation with an external electric field. Their asymptotics as $t<t_{1}$ can
be classified as free particles and antiparticles. The $\mathrm{out}$%
-operators are associated with a complete orthonormal $\mathrm{out}$-set of
solutions $\left\{ \ ^{\zeta }\psi _{n}\left( t,\mathbf{r}\right) \right\} $
of the Dirac equation with an external electric field. Their asymptotics as $%
t>t_{2}$ can be classified as free particles and antiparticles. The
conserved inner product reads%
\begin{equation*}
\left( \psi ,\psi ^{\prime }\right) =\int \psi ^{\dag }\left( t,\mathbf{r}%
\right) \psi ^{\prime }\left( t,\mathbf{r}\right) d\mathbf{r},
\end{equation*}%
where the integration is over the finite area $S$ of the standard box
regularization. The orthonormality conditions are:%
\begin{equation}
\left( \ _{\zeta }\psi _{n},\ _{\zeta ^{\prime }}\psi _{n^{\prime }}\right)
=\delta _{\zeta ,\zeta ^{\prime }}\delta _{n,n^{\prime }},\;\;\left( \
^{\zeta }\psi _{n},\ ^{\zeta ^{\prime }}\psi _{n^{\prime }}\right) =\delta
_{\zeta ,\zeta ^{\prime }}\delta _{n,n^{\prime }}.  \label{norm}
\end{equation}

The $\mathrm{in}$- and $\mathrm{out}$- operators are defined by the two
representations of the quantum Dirac field $\hat{\Psi}\left( t,\mathbf{r}%
\right) $ in the Heisenberg representation (It means here: in the zero-order
approximations with respect of interaction with photons)
\begin{align}
& \ \hat{\Psi}\left( t,\mathbf{r}\right) =\sum_{n}\left[ a_{n}(\mathrm{in}%
)\;_{+}\psi _{n}\left( t,\mathbf{r}\right) +b_{n}^{\dagger }(\mathrm{in}%
)\;_{-}\psi _{n}\left( t,\mathbf{r}\right) \right]  \notag \\
& =\sum_{n}\left[ a_{n}(\mathrm{out})\;^{+}\psi _{n}\left( t,\mathbf{r}%
\right) +b_{n}^{\dagger }(\mathrm{out}))\;^{-}\psi _{n}\left( t,\mathbf{r}%
\right) \right] \ .  \label{f6}
\end{align}%
The $\mathrm{in}$- and $\mathrm{out}$-solutions with given quantum numbers $%
n $ are related by a linear transformation of the form:
\begin{eqnarray}
\ ^{\zeta }\psi _{n}\left( t,\mathbf{r}\right) &=&g_{n}\left( _{+}\left\vert
{}\right. ^{\zeta }\right) \,_{+}\psi _{n}\left( t,\mathbf{r}\right)
+g_{n}\left( _{-}\left\vert {}\right. ^{\zeta }\right) \,_{-}\psi _{n}\left(
t,\mathbf{r}\right) \,,  \notag \\
\ _{\zeta }\psi _{n}\left( t,\mathbf{r}\right) &=&g_{n}\left( ^{+}\left\vert
{}\right. _{\zeta }\right) \,^{+}\psi _{n}\left( t,\mathbf{r}\right)
+g_{n}\left( ^{-}\left\vert {}\right. _{\zeta }\right) \,^{-}\psi _{n}\left(
t,\mathbf{r}\right) \,,  \label{f7}
\end{eqnarray}
where the $g^{\prime }$s are some complex coefficients, $g\left( ^{\zeta
^{\prime }}\left\vert {}\right. _{\zeta }\right) =\left( _{\zeta }\left\vert
{}\right. ^{\zeta ^{\prime }}\right) ^{\ast }$ . These coefficients obey the
unitarity relations:
\begin{align}
& g_{n}\left( ^{\zeta }\left\vert {}\right. _{+}\right) g_{n}\left(
_{+}\left\vert {}\right. ^{\zeta }\right) +g_{n}\left( ^{\zeta }\left\vert
{}\right. _{-}\right) g_{n}\left( _{-}\left\vert {}\right. ^{\zeta }\right)
=1\,,  \notag \\
& g_{n}\left( _{\zeta }\left\vert {}\right. ^{+}\right) g_{n}\left(
^{+}\left\vert {}\right. _{\zeta }\right) +g_{n}\left( _{\zeta }\left\vert
{}\right. ^{-}\right) g_{n}\left( ^{-}\left\vert {}\right. _{\zeta }\right)
=1\,,  \notag \\
& g_{n}\left( _{+}\left\vert {}\right. ^{+}\right) g_{n}\left(
^{+}\left\vert {}\right. _{-}\right) +g_{n}\left( _{+}\left\vert {}\right.
^{-}\right) g_{n}\left( ^{-}\left\vert {}\right. _{-}\right) =0\,,  \notag \\
& g_{n}\left( ^{+}\left\vert {}\right. _{+}\right) g_{n}\left(
_{+}\left\vert {}\right. ^{-}\right) +g_{n}\left( ^{+}\left\vert {}\right.
_{-}\right) g_{n}\left( _{-}\left\vert {}\right. ^{-}\right) =0\,,
\label{f8}
\end{align}
which follow from the orthonormalization and completeness relations for the
corresponding solutions. It is known that all the coefficients can be
expressed in terms of two of them, e.g. of $g\left( _{+}\left\vert
^{+}\right. \right) $ and $g\left( _{-}\left\vert ^{+}\right. \right) $.
However, even the latter coefficients are not completely independent,
\begin{equation}
\left\vert g_{n}\left( _{-}\left\vert ^{+}\right. \right) \right\vert
^{2}+\left\vert g_{n}\left( _{+}\left\vert ^{+}\right. \right) \right\vert
^{2}=1.  \label{f9}
\end{equation}

Then a linear canonical transformation (Bogolubov transformation) between $%
\mathrm{in}$- and $\mathrm{out}$- operators which follows from Eq.~(\ref{f6}%
) is defined by these coefficients
\begin{align}
a_{n}\left( \mathrm{out}\right) & =g_{n}\left( ^{+}\left\vert {}\right.
_{+}\right) a_{n}(\mathrm{in})+g_{n}\left( ^{+}\left\vert {}\right.
_{-}\right) b_{n}^{\dagger }(\mathrm{in}),  \notag \\
b_{n}^{\dagger }\left( \mathrm{out}\right) & =g_{n}\left( ^{-}\left\vert
{}\right. _{+}\right) a_{n}(\mathrm{in})+g_{n}\left( ^{-}\left\vert
{}\right. _{-}\right) b_{n}^{\dagger }(\mathrm{in}).  \label{f10}
\end{align}

Using relations (\ref{f10}), one finds that the differential mean numbers $%
N_{n}^{(\zeta )}$ of electrons (holes) created from the vacuum in the
zero-order approximations with respect of the electron-photon interaction
are:%
\begin{eqnarray}
N_{n}^{(+)} &=&\ _{\mathrm{e}}\left\langle 0,\mathrm{in}\right\vert
a_{n}^{\dagger }(\mathrm{out})a_{n}(\mathrm{out})\left\vert 0,\mathrm{in}%
\right\rangle _{\mathrm{e}}\ =\left\vert g_{n}\left( _{-}\left\vert
{}\right. ^{+}\right) \right\vert ^{2},  \notag \\
N_{n}^{(-)} &=&\ _{\mathrm{e}}\left\langle 0,\mathrm{in}\right\vert
b_{n}^{\dagger }(\mathrm{out})b_{n}(\mathrm{out})\left\vert 0,\mathrm{in}%
\right\rangle _{\mathrm{e}}=\left\vert g_{n}\left( _{+}\left\vert {}\right.
^{-}\right) \right\vert ^{2}.  \label{f14}
\end{eqnarray}%
We see that that the mean numbers of electrons (holes) created are equal and
are also equal to the mean number of the pairs created, $%
N_{n}^{(+)}=N_{n}^{(-)}=N_{n}^{\mathrm{cr}}$. All the information about
electrons and holes creation, annihilation, and scattering in an electric
field in the zero-order approximations with respect of the electron-photon
interaction can be extracted from the coefficients can be extracted from the
coefficients $g\left( _{\zeta }\left\vert {}\right. ^{\zeta ^{\prime
}}\right) $, see Ref. \cite{GenTheory1,GenTheory2,GenTheory3,FGS91} for
details.

\subsection{In- and out-states with definite numbers of charged particles
and photons\label{SS2.3}}

We note that the Fock space of the complete system under consideration is a
tensor product of the Fock space of the electron subsystem and the Fock
space of the free photon subsystem. As was pointed out above due to the
vacuum instability the $\mathrm{in}$- and \textrm{out}-states of the
electron subsystem are different in the general case. At the same time the
photon vacuum $\left\vert 0\right\rangle _{\mathrm{\gamma }}$ remains
unchanged. Denoting by $\left\vert 0,\mathrm{in}\right\rangle $ and $%
\left\vert 0,\mathrm{out}\right\rangle $ the initial and final vacuum states
of the complete system, we can write:%
\begin{equation*}
\left\vert 0,\mathrm{in}\right\rangle =\left\vert 0,\mathrm{in}\right\rangle
_{\mathrm{e}}\ \otimes \left\vert 0\right\rangle _{\mathrm{\gamma }},\
\left\vert 0,\mathrm{out}\right\rangle =\left\vert 0,\mathrm{out}%
\right\rangle _{\mathrm{e}}\otimes \left\vert 0\right\rangle _{\mathrm{%
\gamma }}\ .
\end{equation*}%
The initial and final states of the complete system with definite numbers of
charged particles of all the kinds and photons have the form:

\begin{eqnarray}
&&\ \ \left\vert \mathrm{in}\right\rangle =C^{\dagger }\ldots b^{\dagger
}\left( \mathrm{in}\right) \ldots a^{\dagger }\left( \mathrm{in}\right)
\ldots \left\vert 0,\mathrm{in}\right\rangle \ ,  \notag \\
&&\ \left\vert \mathrm{out}\right\rangle \ =C^{\dagger }\ldots b^{\dagger
}\left( \mathrm{out}\right) \ldots a^{\dagger }\left( \mathrm{out}\right)
\ldots \left\vert 0,\mathrm{out}\right\rangle .  \label{f18}
\end{eqnarray}

Probability amplitude of a transition from an initial to a final state (\ref%
{f18}) has the following form:%
\begin{equation}
W=\ \left\langle \mathrm{out}\right\vert \mathcal{S}\left\vert \mathrm{in}%
\right\rangle ,  \label{f19}
\end{equation}%
where $\mathcal{S}$ is the scattering matrix in the external field,
\begin{eqnarray}
&&\mathcal{S}=\mathcal{T}\exp \left\{ -\frac{i}{\hbar }%
\int_{t_{in}}^{t_{out}}\mathcal{H}_{\mathrm{int}}dt\right\} ,\;\;\mathcal{H}%
_{\mathrm{int}}\approx -\int \mathbf{j(}t\mathbf{,\mathbf{r})A}(t,\mathbf{r}%
)d\mathbf{r\ },  \notag \\
&&\mathbf{j(}t\mathbf{,\mathbf{r})}=-\frac{ev_{F}}{2c}\left[ \Psi ^{\dag
}\left( t,\mathbf{\mathbf{r}}\right) ,\gamma ^{0}\mathbf{\gamma }\Psi \left(
t,\mathbf{\mathbf{r}}\right) \right] _{-}\;,  \label{f20a}
\end{eqnarray}%
where $\Psi \left( t,\mathbf{r}\right) $, $\Psi ^{\dag }\left( t,\mathbf{r}%
\right) $, and $\mathbf{A}(t,\mathbf{r})$ are quantum field operators in the
interaction representation, the symbol $\mathcal{T}$- is the chronological
ordering operator and $t_{out}-t_{in}\rightarrow \infty $ is macroscopic
time of the radiative interaction.

Below we are going to consider the emission of a photon from the vacuum and
from a single-electron state. These processes will be studied in the
first-order approximation for amplitudes which corresponds to the
second-order approximation for the probabilities. In this case:

\begin{eqnarray}
\mathcal{S} &\approx &1+i\Upsilon ^{\left( 1\right) },\; \Upsilon ^{\left(
1\right) }=\frac{1}{\hbar }\int \mathbf{j(}t\mathbf{,\mathbf{r})A}(t,\mathbf{%
r})d\mathbf{r}dt  \notag \\
&\Longrightarrow &\;\left\langle \mathrm{out}\right\vert \mathcal{S}%
\left\vert \mathrm{in}\right\rangle \approx i\left\langle \mathrm{out}%
\right\vert \Upsilon ^{\left( 1\right) }\left\vert \mathrm{in}\right\rangle
\label{f27b}
\end{eqnarray}

It is known that the QED$_{3,2}$\ model is renormalizable.\emph{\ }In the
first-order approximation we only have to believe that fields, the electric
charge, and electron mass (if $m\neq 0$) are already given in the
renormalized form, namely the charge $e$\ represents its physical value and
the fine-structure constant is $\alpha =e^{2}/\hbar c\simeq 1/137$.

The probabilities of the one-photon emissions with quantum numbers $\mathbf{K%
}$, $\vartheta $ and at the same time with the production of $M\geq 1$ pairs
of charged particles of one kind from the vacuum read:%
\begin{eqnarray}
\mathcal{P}_{M}\left( \mathbf{K,}\vartheta \right) &=&\sum_{\{m\}\left\{
n\right\} }\left( M!\right) ^{-2}\left\vert \left\langle 0,\mathrm{out}%
\right\vert b_{n_{M}}\left( \mathrm{out}\right) \ldots b_{n_{1}}\left(
\mathrm{out}\right) \right. \   \notag \\
&&\times \left. a_{m_{M}}\left( \mathrm{out}\right) \ldots a_{m_{1}}\left(
\mathrm{out}\right) C_{\mathbf{K}\vartheta }i\Upsilon ^{\left( 1\right)
}\left\vert 0,\mathrm{in}\right\rangle \right\vert ^{2}.  \label{f30}
\end{eqnarray}%
Summing these probabilities over $M$, we obtain the probability of the
one-photon emission with full allowance for the possible instability of the
vacuum with respect to the production of one kind of charged particles,%
\begin{equation}
\mathcal{P}\left( \mathbf{K,}\vartheta \right) =\sum_{M=1}^{\infty }\mathcal{%
P}_{M}\left( \mathbf{K,}\vartheta \right) .  \label{fPvac}
\end{equation}

The probabilities of the one-photon emission with quantum numbers $\mathbf{K}
$, $\vartheta $ and at the same time with the production of $M\geq 0$ pairs
of charged particles of one kind from a single-electron state which is
characterized by a quantum numbers $l$ reads:
\begin{eqnarray}
&&\ \mathcal{P}_{M}\left( \left. \mathbf{K},\vartheta \right\vert \overset{+}%
{l}\right) =\sum_{\{m\}\left\{ n\right\} }\left[ M!\left( M+1\right) !\right]
^{-1}\left\vert \left\langle 0,\mathrm{out}\right\vert b_{n_{M}}\left(
\mathrm{out}\right) \ldots b_{n_{1}}\left( \mathrm{out}\right) \right.
\notag \\
&&\times \left. a_{m_{M+1}}\left( \mathrm{out}\right) \ldots a_{m_{1}}\left(
\mathrm{out}\right) C_{\mathbf{K}\vartheta }i\Upsilon ^{\left( 1\right)
}a_{l}^{\dagger }(\mathrm{in})\left\vert 0,\mathrm{in}\right\rangle
\right\vert ^{2}.  \label{f32}
\end{eqnarray}%
The same probability from a single-hole state has the form:

\begin{eqnarray}
\mathcal{P}_{M}\left( \left. \mathbf{K},\vartheta \right\vert \overset{-}{l}%
\right) &=&\sum_{\{m\}\left\{ n\right\} }\left[ M!\left( M+1\right) !\right]
^{-1}\left\vert \left\langle 0,\mathrm{out}\right\vert b_{n_{M}}\left(
\mathrm{out}\right) \ldots b_{n_{1}}\left( \mathrm{out}\right) \right.
\notag \\
&&\times \left. a_{m_{M+1}}\left( \mathrm{out}\right) \ldots a_{m_{1}}\left(
\mathrm{out}\right) C_{\mathbf{K}\vartheta }i\Upsilon ^{\left( 1\right)
}b_{l}^{\dagger }(\mathrm{in})\left\vert 0,\mathrm{in}\right\rangle
\right\vert ^{2}.  \label{f32b}
\end{eqnarray}

Summing these probabilities over $M$, we obtain the probability of the
one-photon emission from a single-electron (hole) state with full allowance
for the possible instability of the vacuum with respect to the production of
charged particles of one kind,%
\begin{equation}
\mathcal{P}\left( \left. \mathbf{K},\vartheta \right\vert \overset{\pm }{l}%
\right) =\sum_{M=0}^{\infty }\mathcal{P}_{M}\left( \left. \mathbf{K}%
,\vartheta \right\vert \overset{\pm }{l}\right) .  \label{f40}
\end{equation}

If the probability of the creation of pairs from the vacuum is small then
main contributions to probabilities (\ref{f40}) and (\ref{fPvac}) are due
minimal possible numbers of created pairs,%
\begin{equation}
\mathcal{P}\left( \mathbf{K,}\vartheta \right) \approx \mathcal{P}_{1}\left(
\mathbf{K,}\vartheta \right) ,\;\mathcal{P}\left( \left. \mathbf{K}%
,\vartheta \right\vert \overset{\pm }{l}\right) \approx \mathcal{P}%
_{0}\left( \left. \mathbf{K},\vartheta \right\vert \overset{\pm }{l}\right)
\ .  \label{f41}
\end{equation}

To construct an perturbation theory for the probability amplitudes one needs
to reduce the $S$-matrix to a generalized normal form with respect to the
vacua $\left\langle 0,\mathrm{out}\right\vert $ and $\left\vert 0,\mathrm{in}%
\right\rangle $, see Ref. \cite{GenTheory1,GenTheory2,GenTheory3,FGS91}.

To this end, one has explicitly divide the Dirac field operators into parts,
creative with respect to the vacuum $\left\langle 0,\mathrm{out}\right\vert $
and annihilative with respect to the vacuum $\left\vert 0,\mathrm{in}%
\right\rangle $. In the first-order approximation, it is sufficient to
reduce the operator $\mathbf{j(}t\mathbf{,\mathbf{r})}$ to the generalized
normal form,
\begin{eqnarray}
&&\mathcal{N}_{out-in}\left\{ \mathbf{j(}t\mathbf{,\mathbf{r})}\right\} =\
\mathbf{j(}t\mathbf{,\mathbf{r})}-\langle \mathbf{j(}t\mathbf{,\mathbf{r})}%
\rangle ^{c},  \notag \\
&&\langle \mathbf{j(}t\mathbf{,\mathbf{r})}\rangle ^{c}=\left\langle 0,%
\mathrm{out}\right\vert \mathbf{j(}t\mathbf{,\mathbf{r})}\left\vert 0,%
\mathrm{in}\right\rangle c_{v}^{-1},  \label{f26}
\end{eqnarray}%
where $c_{v}=\left\langle 0,\mathrm{out}\right. \left\vert 0,\mathrm{in}%
\right\rangle $\ is the vacuum to vacuum transition amplitude. Note that, in
the general case, the vacuum polarization current $\langle \mathbf{j(}t%
\mathbf{,\mathbf{r})}\rangle ^{c}$ is not zero. It may contribute to a
tadpole{\Huge \ }diagram.\emph{\ }However, in the case under consideration
(uniform and slowly varying external electric field),\emph{\ }such a diagram
has a notable value only in a very infrared range, which is not considered
here.\emph{\ }In particular, in the limiting case of a constant uniform
field, this contribution can be safely neglected\footnote{%
We note that this diagram can cause non-vanishing contributions when
appearing as a part of a higer order diagram; see Ref.\ \cite{GiesK17}}.

Thus, the quantities under consideration, can be represented as:%
\begin{eqnarray}
&&\mathcal{P}_{0}\left( \left. \mathbf{K},\vartheta \right\vert \overset{\pm
}{l}\right) =P_{v}^{\left( 0\right) }\sum_{n}\left\vert w^{\left( 1\right)
}\left( \overset{\pm }{n};\left. \mathbf{K,}\vartheta \right\vert \overset{%
\pm }{l}\right) \right\vert ^{2},\;P_{v}^{\left( 0\right) }=\left\vert
c_{v}\right\vert ^{2},\   \notag \\
&&w^{\left( 1\right) }\left( \overset{\pm }{n};\left. \mathbf{K,}\vartheta
\right\vert \overset{\pm }{l}\right) =\frac{ic}{\hbar }\sqrt{\frac{2\pi
\hbar }{\varepsilon V\omega }}\int \boldsymbol{\epsilon }_{\mathbf{K}%
\vartheta }\mathbf{j}\left( \left. \overset{\pm }{n}\right\vert \overset{\pm
}{l}\right) e^{i(\omega t-\mathbf{kr})}dtd\mathbf{r\ },  \notag \\
&&\mathbf{j}\left( \left. \overset{+}{n}\right\vert \overset{+}{l}\right)
=c_{v}^{-1}\left\langle 0,\mathrm{out}\right\vert a_{n}\left( \mathrm{out}%
\right) \mathcal{N}_{out-in}\left\{ \mathbf{j(}t\mathbf{,\mathbf{r})}%
\right\} a_{l}^{\dagger }(\mathrm{in})\left\vert 0,\mathrm{in}\right\rangle ,
\notag \\
&&\mathbf{j}\left( \left. \overset{-}{n}\right\vert \overset{-}{l}\right)
=c_{v}^{-1}\left\langle 0,\mathrm{out}\right\vert b_{n}\left( \mathrm{out}%
\right) \mathcal{N}_{out-in}\left\{ \mathbf{j(}t\mathbf{,\mathbf{r})}%
\right\} b_{l}^{\dagger }(\mathrm{in})\left\vert 0,\mathrm{in}\right\rangle
\ ,  \label{f28}
\end{eqnarray}%
and%
\begin{eqnarray}
&&\mathcal{P}_{1}\left( \mathbf{K,}\vartheta \right) =P_{v}^{\left( 0\right)
}\sum_{n,l}\left\vert w^{\left( 1\right) }\left( \overset{-}{n}\overset{+}{l}%
;\left. \mathbf{K,}\vartheta \right\vert 0\right) \right\vert ^{2},  \notag
\\
&&w^{\left( 1\right) }\left( \overset{-}{n}\overset{+}{,l};\left. \mathbf{K,}%
\vartheta \right\vert 0\right) =\frac{ic}{\hbar }\sqrt{\frac{2\pi \hbar }{%
\varepsilon V\omega }}\int \boldsymbol{\epsilon }_{\mathbf{K}\vartheta }%
\mathbf{j}\left( \left. \overset{-}{n}\overset{+}{,l}\right\vert 0\right)
e^{i(\omega t-\mathbf{kr})}dtd\mathbf{r\ },  \notag \\
&&\mathbf{j}\left( \left. \overset{-}{n}\overset{+}{,l}\right\vert 0\right)
=c_{v}^{-1}\left\langle 0,\mathrm{out}\right\vert b_{n}(\mathrm{out}%
)a_{l}\left( \mathrm{out}\right) \mathcal{N}_{out-in}\left\{ \mathbf{j(}t%
\mathbf{,\mathbf{r})}\right\} \left\vert 0,\mathrm{in}\right\rangle \ .
\label{f27}
\end{eqnarray}

One can express matrix elements in Eqs.~(\ref{f27}) and (\ref{f28}) via the
solutions$\ _{\zeta }\psi _{n}\left( t,\mathbf{r}\right) $ and $\ ^{\zeta
}\psi _{n}\left( t,\mathbf{r}\right) $, and coefficients $g$ as follows:
\begin{eqnarray}
&&\mathbf{j}\left( \left. \overset{+}{n,}\overset{-}{l}\right\vert 0\right)
=-\frac{ev_{F}}{c}g_{l}\left( _{+}\left\vert {}\right. ^{+}\right)
^{-1}\;_{+}\bar{\psi}_{l}\left( t,\mathbf{r}\right) \mathbf{\gamma }\
_{-}\psi _{n}\left( t,\mathbf{r}\right) g_{n}\left( ^{-}\left\vert {}\right.
_{-}\right) ^{-1},  \notag \\
&&\mathbf{j}\left( \left. \overset{+}{n}\right\vert l^{+}\right) =-\frac{%
ev_{F}}{c}g_{n}\left( _{+}\left\vert {}\right. ^{+}\right) ^{-1}\;_{+}\bar{%
\psi}_{n}\left( t,\mathbf{r}\right) \mathbf{\gamma }\;^{+}\psi _{l}\left( t,%
\mathbf{r}\right) g_{l}\left( _{+}\left\vert {}\right. ^{+}\right) ^{-1},
\notag \\
&&\mathbf{j}\left( \left. \overset{-}{n}\right\vert l^{-}\right) =\frac{%
ev_{F}}{c}g_{l}\left( ^{-}\left\vert {}\right. _{-}\right) ^{-1}\;^{-}\bar{%
\psi}_{l}\left( t,\mathbf{r}\right) \mathbf{\gamma }\;_{-}\psi _{n}\left( t,%
\mathbf{r}\right) g_{n}\left( ^{-}\left\vert {}\right. _{-}\right) ^{-1}\ ,
\label{f29}
\end{eqnarray}%
where $\bar{\psi}_{n}=$ $\psi _{n}^{\dag }\gamma ^{0}$ .

It seems that the matrix elements in Eqs.~(\ref{f30})-(\ref{f40}) can be
written in a similar manner. However, this is only useful in the case of a
not very strong electric field, when the approximation (\ref{f41}) is
applicable. In the case of an intense external field, there exist many
transition channels corresponding to the violation of the vacuum stability.\
Considering the photon emission by massless charged particles in the
graphene any quasiconstant electric field has to be treated as a strong
one.\ By this reason it is effective to calculate mean characteristics of
the emission using the unitarity condition for the $S$\ matrix as in the way
described below.

Probabilities (\ref{fPvac}) and (\ref{f40}) can be represented as a trace of
the operators $C_{\mathbf{K}\vartheta }S\left\vert \mathrm{in}\right\rangle
\left\langle \mathrm{in}\right\vert S^{-1}C_{\mathbf{K}\vartheta }^{\dag }$
with respect to the final basis,
\begin{equation}
\mathcal{P}\left( \left. \mathbf{K},\vartheta \right\vert \mathrm{in}\right)
=\mathrm{tr\,}\left[ C_{\mathbf{K}\vartheta }\mathcal{S}\left\vert \mathrm{in%
}\right\rangle \left\langle \mathrm{in}\right\vert \mathcal{S}^{-1}C_{%
\mathbf{K}\vartheta }^{\dag }\right] \ ,  \label{f42b}
\end{equation}%
where $\left\vert \mathrm{in}\right\rangle $ is one of the following states:
$\left\vert 0,\mathrm{in}\right\rangle $, $a_{l}^{\dagger }(\mathrm{in}%
)\left\vert 0,\mathrm{in}\right\rangle $, or $b_{l}^{\dagger }(\mathrm{in}%
)\left\vert 0,\mathrm{in}\right\rangle $.\ One can see that trace (\ref{f42b}%
) can be written as a mean value of the photon number operator,%
\begin{equation}
\mathcal{P}\left( \left. \mathbf{K},\vartheta \right\vert \mathrm{in}\right)
=\left\langle \mathrm{in}\right\vert \mathcal{S}^{-1}C_{\mathbf{k}\vartheta
}^{\dag }C_{\mathbf{k}\vartheta }\mathcal{S}\left\vert \mathrm{in}%
\right\rangle \ .  \label{ph_number}
\end{equation}

In course of constructing a perturbation theory with respect to the
radiative interaction{\Huge \ }one needs to reorganize the $S$-matrix in the
normal form $:\ldots :$ with respect to the \textrm{in}-vacuum, see Ref.
\cite{GenTheory1,GenTheory2,GenTheory3,FGS91}. In the first-order
approximation, it is sufficient to represent only the operator $\mathbf{j(}t%
\mathbf{,\mathbf{r})}$ in such a form,%
\begin{equation}
\mathbf{j(}t\mathbf{,\mathbf{r})}=\ :\mathbf{j(}t\mathbf{,\mathbf{r})}:+\
\langle \mathbf{j(}t\mathbf{,\mathbf{r})}\rangle _{\mathrm{in}}\ ,\ \
\langle \mathbf{j(}t\mathbf{,\mathbf{r})}\rangle _{\mathrm{in}}\
=\left\langle 0,\mathrm{in}\right\vert \mathbf{j(}t\mathbf{,\mathbf{r})}%
\left\vert 0,\mathrm{in}\right\rangle \ .  \label{in-currentA}
\end{equation}%
The vacuum mean current $\langle \mathbf{j(}t\mathbf{,\mathbf{r})}\rangle _{%
\mathrm{in}}$ is a sum of a vacuum polarization current and of a current of
created charged particles. It is not zero in a slowly varying electric field
and depends on the definition of the initial vacuum, $\left\vert 0,\mathrm{in%
}\right\rangle $ and on the evolution of the electric field from the initial
time\emph{\ }$t_{1}$\emph{\ }of switching on\emph{\ }to the time instant $t$%
\emph{.} After the time $t_{2}$ of switching the electric field off, the
term $\langle \mathbf{j(}t\mathbf{,\mathbf{r})}\rangle _{\mathrm{in}}$
represents the current density of the created pairs of charged particles.
This current is a source in the Maxwell equations for a mean electromagnetic
field. Such a mean field is a slowly varying crossed field emitted
perpendicular to the graphene plane, see Ref. ~\cite{GavGitY12} for details.
One can see that in the frequency range of the photon emission $\omega \gg
T^{-1},$ which is interesting to us, the contribution due to the current $%
\langle \mathbf{j(}t\mathbf{,\mathbf{r})}\rangle _{\mathrm{in}}$ can be
neglected.

In particular, it follows from Eq. (\ref{ph_number}) that probabilities (\ref%
{fPvac}) and (\ref{f40}) read:%
\begin{eqnarray}
&&\mathcal{P}\left( \mathbf{K,}\vartheta \right) =\sum_{l}\mathcal{P}\left(
l;\left. \mathbf{K,}\vartheta \right\vert 0\right) ,\;\mathcal{P}\left(
l;\left. \mathbf{K,}\vartheta \right\vert 0\right) =\sum_{n}\left\vert w_{%
\mathrm{in}}^{\left( 1\right) }\left( \overset{-}{n}\overset{+}{l};\left.
\mathbf{K,}\vartheta \right\vert 0\right) \right\vert ^{2}\ ,  \notag \\
&&w_{\mathrm{in}}^{\left( 1\right) }\left( \overset{-}{n}\overset{+}{,l}%
;\left. \mathbf{K,}\vartheta \right\vert 0\right) =\frac{ic}{\hbar }\sqrt{%
\frac{2\pi \hbar }{\varepsilon V\omega }}\int \boldsymbol{\epsilon }_{%
\mathbf{K}\vartheta }\mathbf{j}_{\mathrm{in}}\left( \left. \overset{-}{n}%
\overset{+}{,l}\right\vert 0\right) e^{i(\omega t-\mathbf{kr})}dtd\mathbf{r\
},  \notag \\
&&\mathbf{j}_{\mathrm{in}}\left( \left. \overset{-}{n}\overset{+}{,l}%
\right\vert 0\right) =\left\langle 0,\mathrm{in}\right\vert b_{n}(\mathrm{in}%
)a_{l}\left( \mathrm{in}\right) :\mathbf{j(}t\mathbf{,\mathbf{r})}%
:\left\vert 0,\mathrm{in}\right\rangle \ ,  \label{f43b}
\end{eqnarray}%
where $\mathcal{P}\left( l;\left. \mathbf{K,}\vartheta \right\vert 0\right) $
is the probability of one photon emission with quantum numbers $\mathbf{K}$,
$\vartheta $ which is accompanied by the production of pairs of one kind
with a quantum number $l$ , and%
\begin{eqnarray}
&&\mathcal{P}\left( \left. \mathbf{K},\vartheta \right\vert \overset{\pm }{l}%
\right) =\sum_{n}\left\vert w_{\mathrm{in}}^{\left( 1\right) }\left( \overset%
{\pm }{n};\left. \mathbf{K},\vartheta \right\vert \overset{\pm }{l}\right)
\right\vert ^{2},  \notag \\
&&w_{\mathrm{in}}^{\left( 1\right) }\left( \overset{\pm }{n};\left. \mathbf{%
K,}\vartheta \right\vert \overset{\pm }{l}\right) =\frac{ic}{\hbar }\sqrt{%
\frac{2\pi \hbar }{\varepsilon V\omega }}\int \boldsymbol{\epsilon }_{%
\mathbf{K}\vartheta }\mathbf{j}_{\mathrm{in}}\left( \left. \overset{\pm }{n}%
\right\vert \overset{\pm }{l}\right) e^{i(\omega t-\mathbf{kr})}dtd\mathbf{r}%
,  \notag \\
&&\mathbf{j}_{\mathrm{in}}\left( \left. \overset{+}{n}\right\vert \overset{+}%
{l}\right) =\left\langle 0,\mathrm{in}\right\vert a_{n}\left( \mathrm{in}%
\right) :\mathbf{j(}t\mathbf{,\mathbf{r})}:a_{l}^{\dagger }(\mathrm{in}%
)\left\vert 0,\mathrm{in}\right\rangle ,  \notag \\
&&\mathbf{j}_{\mathrm{in}}\left( \left. \overset{-}{n}\right\vert \overset{-}%
{l}\right) =\left\langle 0,\mathrm{in}\right\vert b_{n}\left( \mathrm{in}%
\right) :\mathbf{j(}t\mathbf{,\mathbf{r})}:b_{l}^{\dagger }(\mathrm{in}%
)\left\vert 0,\mathrm{in}\right\rangle .  \label{f43c}
\end{eqnarray}%
In order to find the probability of one photon emission with quantum numbers
$\mathbf{K}$, $\vartheta $ which is accompanied by the production of pairs
of all the kinds in the graphene,{\large \ }the probability (\ref{f43b}) is
multiplied by the number of species $N_{f}=4$,%
\begin{equation}
\mathcal{P}_{N_{f}}\left( \mathbf{K,}\vartheta \right) =N_{f}\mathcal{P}%
\left( \mathbf{K,}\vartheta \right) .  \label{f43d}
\end{equation}

One can express matrix elements in Eqs.~(\ref{f43b}) and (\ref{f43c}) via
the solutions $\ _{\zeta }\psi _{n}\left( t,\mathbf{r}\right) $ as follows:
\begin{eqnarray}
&&\mathbf{j}_{\mathrm{in}}\left( \left. \overset{-}{n}\overset{+}{,l}%
\right\vert 0\right) =-\frac{ev_{F}}{c}\;_{+}\bar{\psi}_{l}\left( t,\mathbf{r%
}\right) \mathbf{\gamma }\;_{-}\psi _{n}\left( t,\mathbf{r}\right) ,  \notag
\\
&&\mathbf{j}_{\mathrm{in}}\left( \left. \overset{\pm }{n}\right\vert \overset%
{\pm }{l}\right) =\mp \frac{ev_{F}}{c}\;_{\pm }\bar{\psi}_{n}\left( t,%
\mathbf{r}\right) \mathbf{\gamma }\;_{\pm }\psi _{l}\left( t,\mathbf{r}%
\right) .  \label{f44}
\end{eqnarray}

Note that, in the general case, the matrix elements $w_{\mathrm{in}}^{\left(
1\right) }\left( \overset{-}{n}\overset{+}{,l};\left. \mathbf{K,}\vartheta
\right\vert 0\right) $ and $w_{\mathrm{in}}^{\left( 1\right) }\left( \overset%
{\pm }{n};\left. \mathbf{K,}\vartheta \right\vert \overset{\pm }{l}\right) $
considered separately are quite different from the amplitudes of the
relative probabilities $w^{\left( 1\right) }\left( \overset{-}{n}\overset{+}{%
,l};\left. \mathbf{K,}\vartheta \right\vert 0\right) $ and $w^{\left(
1\right) }\left( \overset{\pm }{n};\left. \mathbf{K,}\vartheta \right\vert
\overset{\pm }{l}\right) $ given by Eqs.~(\ref{f27}) and (\ref{f28}),
respectively. Only if the mean number of the pairs created are sufficiently
small, $N_{n},N_{l}\ll 1$, the difference between the solutions $\ _{\zeta
}\psi _{n}\left( t,\mathbf{r}\right) $ and $\ ^{\zeta }\psi _{n}\left( t,%
\mathbf{r}\right) $ at a given $\zeta $ can be neglected.

\section{Photon emission in a constant electric field\label{S3}}

\subsection{Exact solutions\label{SS3.1}}

Next, we proceed to direct calculations of the photon emission in graphene
induced by applied external electric field. The electric field acts in the
graphene plane during the time interval $t_{2}-t_{1}=T$ as a constant field $%
E$ and vanishes out the interval. Such a field is often called $T$-constant
electric field. In the QED$_{3,2}$ model which describes massless charged
particles, even a seemingly weak electric field $E$, if it remains unchanged
for a sufficiently long time, creates electron-hole pairs from the vacuum.
The vacuum instability problem in the graphene in $T$-constant electric
field was studied in detail in Ref. \cite{GavGitY12}. Switching on and off
effects of in the latter field can be neglected if\ we suppose that the time
interval $T$ is sufficiently large, namely
\begin{equation}
T/\Delta t_{st}\gg \,\max \left\{ 1,\frac{m^{2}v_{F}^{3}}{\left\vert
eE\right\vert \hbar }\right\} \,,  \label{time-condition}
\end{equation}%
where $\Delta t_{st}$ is a big characteristic time scale $\Delta t_{st}$ in
the graphene physics,
\begin{equation}
\Delta t_{st}=\left( \left\vert eE\right\vert v_{F}/\hbar \right)
^{-1/2}\,\gg t_{\gamma }\ .  \label{e1}
\end{equation}%
and $t_{\gamma }=\hbar /\gamma \simeq 0.24\mathrm{fs}$ is the microscopic
time scale with $\gamma =2.7$\textrm{eV} being the hopping\textrm{\ }energy.
Then the perturbation theory with respect to electric field breaks down and
the dc response changes from the linear in $E$ time-independent\emph{\ }
regime to a nonlinear in $E$ and time-dependent regime, see Ref. \cite%
{lewkowicz-10}. This regime was recently observed in measurements of $I-V$
curves of graphene devices near the Dirac point, see Ref. \cite%
{vandecasteele}.

We recall that the $T$-constant electric field can be described by the
vector potential with only one nonzero component $A_{x}^{\mathrm{ext}}(t),$
\begin{equation*}
A_{x}^{\mathrm{ext}}(t)=-cE\left\{
\begin{array}{ll}
t_{1} & t\in \mathrm{I}=(-\infty ,t_{in}),\ t_{1}=-T/2\, \\
t, & t\in \mathrm{Int}=[t_{1},t_{2}]\, \\
t_{2}, & t\in \mathrm{II}=(t_{2},\infty )\,,\ t_{2}=T/2\,%
\end{array}%
\right. ,
\end{equation*}%
such that the corresponding electric field has also only one nonzero
component, $E_{x}\left( t\right) ,$%
\begin{equation*}
E_{x}\left( t\right) =\left\{
\begin{array}{l}
E>0,\ t\in \mathrm{Int} \\
0,\ t\in \mathrm{I}\cup \mathrm{II}%
\end{array}%
\right. .
\end{equation*}

The time scale (\ref{e1}) is specific to the graphene physics. It plays the
role of a stabilization time after which differential mean numbers of
created pairs take the form:
\begin{equation}
N_{\mathbf{p}}^{\mathrm{cr}}\simeq e^{-\pi \lambda }\,,\;\;\lambda =\frac{%
v_{F}p_{y}^{2}+m^{2}v_{F}^{3}}{eE\hbar }.  \label{e2}
\end{equation}%
which is the same for the case of the constant electric field in the finite
momentum range%
\begin{equation}
D:\sqrt{\frac{v_{F}}{eE\hbar }}\left\vert p_{y}\right\vert <\left( T/\Delta
t_{st}-\tau \right) ^{1/2},\;\sqrt{\frac{v_{F}}{eE\hbar }}\left\vert
p_{x}\right\vert <\frac{1}{2}T/\Delta t_{st}-\tau ,  \label{e3}
\end{equation}%
where $\tau $ is an arbitrary number satisfying the condition%
\begin{equation}
T/\Delta t_{st}\gg \tau \gg \,\max \left\{ 1,\frac{m^{2}v_{F}^{3}}{eE\hbar }%
\right\} ;  \label{Tcond}
\end{equation}%
see Ref. \cite{GavGitY12} for details. The total number density of
electron-hole pairs created by the electric field (multiplied by a
degeneracy factor $N_{f}=4$) is:%
\begin{equation}
n_{g}^{cr}=r_{g}^{cr}T\,,\quad r_{g}^{cr}=N_{f}\left( 2\pi \right)
^{-2}\left( v_{F}\hbar ^{3}\right) ^{-1/2}\left\vert eE\right\vert ^{3/2}\,.
\label{n-cr}
\end{equation}

The QED with an external constant electric field is a consistent model as
long as the low-frequency radiation field produced by an induced current is
negligible compared to the external field, which is supported by external
sources to remain fixed. In the graphene with zero mass carries this imposes
the consistency restriction \cite{GavGitY12}:
\begin{equation}
\,T/\Delta t_{st}\ll \pi /4\alpha \,,  \label{g8}
\end{equation}%
where $\alpha $ is the fine-structure constant.

In this case under consideration, the external field can be considered as a
good approximation of the effective mean field. In the presence of the mass
gap $\Delta \varepsilon =mv_{F}^{2}$, {restriction }(\ref{g8}) is attenuated
by a factor $\exp \left[ \pi (\Delta \varepsilon )^{2}/e\left\vert
E\right\vert v_{F}\hbar \right] $; see Ref. \cite{GG08-a}. We call the
typical time scale related to Eq.~(\ref{g8}), $\Delta t_{br}=\Delta
t_{st}\pi /4\alpha $, the time of backreaction. On the other hand, the
dimensionless parameter in the lhs of Eq.~(\ref{g8}) satisfies the condition
given by Eq.~(\ref{time-condition}). Thus, there is a {window} in the
parameter range where the model is consistent, $t_{\gamma }\ll \Delta
t_{st}\ll \Delta t_{br}$. Moreover, this restriction corresponds to a
specific regime, which might be relevant to some known experiments in the
graphene physics, see \cite{GavGitY12,emis-exp17}{\ for details}. The time $%
T $ could be treated as a typical time-scale, which we call the effective
time duration $T_{eff}$, $T=T_{eff}$ in what follows. Some kind of
dissipation process may truncate the motion of a particle at $T_{dis}$, in
which case $T_{eff}=T_{dis}$. In the absence of the dissipation, the
transport is ballistic; then, considering a strip with a lateral infinite
width and a finite length $L_{x}$, we assume the ballistic flight time $%
T_{bal}=L_{x}/v_{F}$ to be the effective time duration, $T_{eff}=T_{bal}$.
In typical experiments, $L_{x}\sim 1\mathrm{\mu m}$, which gives $%
T_{bal}\sim 10^{-12}\mathrm{s}$. We note that the experimentally terahertz
pulses \cite{emis-exp17} are characterized by a similar to $T_{bal}$ period.
Taking $T=T_{bal}$ in Eqs.~(\ref{Tcond}) and (\ref{g8}), we obtain the
following restrictions on the constant electric field under consideration:
\begin{equation}
E=aE_{0},\;\;E_{0}=1\times 10^{6}\mathrm{V/m},\;\;7\times 10^{-4}\ll a\ll 8.
\label{g9}
\end{equation}%
Since the voltage is $V=EL_{x}$, one finds the inequalities
\begin{equation}
7\times 10^{-4}\,\mathrm{V}\ll V\ll 8\,\mathrm{V}\,.
\label{eq:voltage-bounds}
\end{equation}%
These voltages are in the range used in experiments in graphene physics.

Solutions of the Dirac equation (\ref{Dext}) with the $T$-constant field
were studied in details in Ref.~\cite{GG96}. It was demonstrated that the
corresponding initial set $\left\{ _{\zeta }\psi _{n}\left( t,\mathbf{r}%
\right) \right\} $ and final set $\left\{ ^{\zeta }\psi _{n}\left( t,\mathbf{%
r}\right) \right\} $ can be chosen in the form:
\begin{gather}
_{\pm }\psi _{\mathbf{p}}(t,\mathbf{r})=(i\hbar \,\partial _{t}+H^{\mathrm{%
ext}})\;_{\pm }\phi _{\mathbf{p},\pm 1}(t,\mathbf{r})\,,\;\;_{\pm }\phi _{%
\mathbf{p},\pm 1}(t,\mathbf{r})=e^{i\mathbf{p}\cdot \mathbf{r}/\hbar
}\;_{\pm }\varphi _{\mathbf{p},\pm 1}(t)U_{\pm 1},  \notag \\
^{\pm }\psi _{\mathbf{p}}(t,\mathbf{r})=(i\hbar \,\partial _{t}+H^{\mathrm{%
ext}})\;^{\pm }\phi _{\mathbf{p},\mp 1}(t,\mathbf{r})\,,\;\;^{\pm }\phi _{%
\mathbf{p},\mp }(t,\mathbf{r})=e^{i\mathbf{p}\cdot \mathbf{r}/\hbar }\;^{\pm
}\varphi _{\mathbf{p},\mp 1}(t)U_{\mp 1}\,,  \label{e4}
\end{gather}%
where $U_{s}$ are constant orthonormalized spinors
\begin{equation*}
U_{+1}=\frac{1}{\sqrt{2}}\left(
\begin{array}{c}
1 \\
1%
\end{array}%
\right) \,,\ U_{-1}=\frac{1}{\sqrt{2}}\left(
\begin{array}{c}
1 \\
-1%
\end{array}%
\right) \,.
\end{equation*}%
At early ($t<t_{1}$ -region $\mathrm{I}$) and late ($t>t_{2}$ -region $%
\mathrm{II}$) times, Eq.~(\ref{Dext}) has plane wave solutions $_{\pm
}\varphi _{\mathbf{p},s}(t)$ and $^{\pm }\varphi _{\mathbf{p},s}(t)$,
respectively, which satisfy simple dispersion relations:%
\begin{align}
& \mathrm{I}:\ _{\zeta }\varphi _{\mathbf{p},s}(t)\sim e^{-i\zeta
\varepsilon _{in}t/\hbar }\,,\quad \mathrm{II}:\ ^{\zeta }\varphi _{\mathbf{p%
},s}(t)\sim e^{-i\zeta \varepsilon _{out}t/\hbar }\,,  \notag \\
& \varepsilon _{in/out}=v_{F}\sqrt{%
(p_{x}-eEt_{1/2})^{2}+p_{y}^{2}+m^{2}v_{F}^{2}}\,.  \label{e5}
\end{align}

For $t\in \mathrm{Int}$, if the electric field satisfies condition (\ref%
{time-condition}), it is enough to use solutions (\ref{e4}) with momenta in
range (\ref{e3}). In this range of momenta, the function $_{\pm }\varphi _{%
\mathbf{p},s}(t)$ and $^{\pm }\varphi _{\mathbf{p},s}(t)$ have the form of
the Weber parabolic cylinder functions (WPCF's):%
\begin{eqnarray}
&&_{+}^{-}\varphi _{\mathbf{p},s}(t)=CD_{\nu -\frac{1+s}{2}}[\pm (1-i)\xi
]\,,_{-}^{+}\varphi _{\mathbf{p},s}(t)=CD_{-\nu -\frac{1-s}{2}}[\pm (1+i)\xi
]\,,  \notag \\
&&\xi =\sqrt{\frac{v_{F}}{eE\hbar }}\left( eEt-p_{x}\right) \,,\ \nu =\frac{%
i\lambda }{2}\,,\ C=\left( 2eE\hbar v_{F}S\right) ^{-1/2}\exp \left( -\frac{%
\pi \lambda }{8}\right) ,  \label{WPCa}
\end{eqnarray}%
where $S$ is the graphene area. An initial state $_{\pm }\psi _{\mathbf{p}%
}(t,\mathbf{r})$ describes a particle/hole with a well-defined energy $%
\varepsilon _{in}$ at the distant past. Similarly, a final state $^{\pm
}\psi _{\mathbf{p}}(t,\mathbf{r})$ describes a particle/hole with a
well-defined energy $\varepsilon _{out}$ at the distant future. Then, the
probability of the emission of a photon in the $T$-constant electric field
during the time interval $T${\Huge \ }is indistinguishable from the one in
the constant field ($T\rightarrow \infty $). Therefore, we assume in what
follows that $T\rightarrow \infty $.

We note that calculating amplitudes (\ref{f44}), it is convenient to
represent solutions $_{\pm }\psi _{\mathbf{p}}(t,\mathbf{r})$ in a different
form. By inserting Eq.~(\ref{WPCa}) in Eq.~(\ref{e4}), and taking explicitly
derivatives, we find this form:%
\begin{eqnarray}
&&_{\pm }\psi _{\mathbf{p}}(t,\mathbf{r})=e^{i\mathbf{pr}/\hbar }\;_{\pm
}\psi _{\mathbf{p}}(t),\ \   \notag \\
&&_{\pm }\psi _{\mathbf{p}}(t)=v_{F}\left[ (mv_{F}\mp i\zeta p_{y})\;_{\pm
}\varphi _{\mathbf{p},\pm 1}(t)U_{\mp 1}+(\pm 1+i)\sqrt{\frac{eE\hbar }{v_{F}%
}}\;_{\pm }\varphi _{\mathbf{p},\mp 1}(t)U_{\pm 1}\right] .  \label{e6}
\end{eqnarray}

\subsection{Probabilities\label{SS3.2}}

The differential probability of one photon emission with a given
polarization $\vartheta $ and the wave vector situated in the range from $%
\mathbf{K}$ to $\mathbf{K+}d\mathbf{K}$\textbf{,} which is accompanied by
the pair production of one kind from the vacuum, reads:
\begin{subequations}
\begin{equation}
d\mathcal{P}\left( \mathbf{K,}\vartheta \right) =\mathcal{P}_{N_{f}}\left(
\mathbf{K,}\vartheta \right) \frac{Vd\mathbf{K}}{\left( 2\pi \right) ^{3}},
\label{e7a2}
\end{equation}%
where the quantity $\mathcal{P}_{N_{f}}\left( \mathbf{K,}\vartheta \right) $
is given by Eqs. (\ref{f43b}) and (\ref{f43d}).

The differential probability of one photon emission with a given
polarization $\vartheta $ and a wave vector situated in the range from $%
\mathbf{K}$ to $\mathbf{K+}d\mathbf{K}$ from a single-electron (hole) state
is
\end{subequations}
\begin{equation}
d\mathcal{P}\left( \left. \mathbf{K,}\vartheta \right\vert \overset{\pm }{%
\mathbf{p}}\right) =\mathcal{P}\left( \left. \mathbf{K,}\vartheta
\right\vert \overset{\pm }{\mathbf{p}}\right) \frac{Vd\mathbf{K}}{\left(
2\pi \right) ^{3}},  \label{e7a1}
\end{equation}%
where the probability $\mathcal{P}\left( \left. \mathbf{K,}\vartheta
\right\vert \overset{\pm }{\mathbf{p}}\right) $ is given by Eq. (\ref{f43c}).

Using the parametrization by frequency $\omega $ and solid angle $d\Omega $,
$d\mathbf{K=}c^{-3}\omega ^{2}d\omega d\Omega $, one can write the
probabilities per unit frequency and solid angle as%
\begin{eqnarray}
&&\frac{d\mathcal{P}\left( \mathbf{K,}\vartheta \right) }{d\omega d\Omega }%
=N_{f}\sum_{\mathbf{p}}\frac{d\mathcal{P}\left( \mathbf{p};\left. \mathbf{K,}%
\vartheta \right\vert 0\right) }{d\omega d\Omega },\;  \notag \\
&&\frac{d\mathcal{P}\left( \mathbf{p};\left. \mathbf{K,}\vartheta
\right\vert 0\right) }{d\omega d\Omega }=\frac{V\omega ^{2}}{\left( 2\pi
c\right) ^{3}}\sum_{\mathbf{p}^{\prime }}\left\vert w_{\mathrm{in}}^{\left(
1\right) }\left( \overset{-}{\mathbf{p}}\overset{+}{\mathbf{p}^{\prime }}%
;\left. \mathbf{K,}\vartheta \right\vert 0\right) \right\vert ^{2};
\label{e7a3} \\
&&\frac{d\mathcal{P}\left( \left. \mathbf{K,}\vartheta \right\vert \overset{%
\pm }{\mathbf{p}}\right) }{d\omega d\Omega }=\frac{V\omega ^{2}}{\left( 2\pi
c\right) ^{3}}\sum_{\mathbf{p}^{\prime }}\left\vert w_{\mathrm{in}}^{\left(
1\right) }\left( \overset{\pm }{\mathbf{p}^{\prime }};\left. \mathbf{K,}%
\vartheta \right\vert \overset{\pm }{\mathbf{p}}\right) \right\vert ^{2},
\label{e7b}
\end{eqnarray}%
where the amplitudes are given by Eqs.~(\ref{f43b}) and (\ref{f43c}),
respectively. Integrating over the area $S$ we obtain that%
\begin{eqnarray}
&&w_{\mathrm{in}}^{\left( 1\right) }\left( \overset{-}{\mathbf{p}}\overset{+}%
{\mathbf{p}^{\prime }};\left. \mathbf{K,}\vartheta \right\vert 0\right)
=iev_{F}\Delta t_{st}\sqrt{\frac{2\pi }{\hbar \varepsilon V\omega }}\delta _{%
\mathbf{p}^{\prime },\mathbf{p}-\hbar \mathbf{k}}M_{\mathbf{p}^{\prime }%
\mathbf{p}}^{0}\ ,  \notag \\
&&M_{\mathbf{p}^{\prime }\mathbf{p}}^{0}=-\frac{S}{\Delta t_{st}}%
\int_{t_{1}}^{t_{2}}\;_{+}\bar{\psi}_{\mathbf{p}^{\prime }}(t)\mathbf{\gamma
}\boldsymbol{\epsilon }_{\mathbf{K}\vartheta }\;_{-}\psi _{\mathbf{p}%
}(t)e^{i\omega t}dt\ ;  \notag \\
&&w_{\mathrm{in}}^{\left( 1\right) }\left( \overset{\pm }{\mathbf{p}^{\prime
}};\left. \mathbf{K,}\vartheta \right\vert \overset{\pm }{\mathbf{p}}\right)
=iev_{F}\Delta t_{st}\sqrt{\frac{2\pi }{\hbar \varepsilon V\omega }}\delta _{%
\mathbf{p}^{\prime },\mathbf{p}-\hbar \mathbf{k}}M_{\mathbf{p}^{\prime }%
\mathbf{p}}^{\pm }\ ,  \notag \\
&&M_{\mathbf{p}^{\prime }\mathbf{p}}^{\pm }=\mp \frac{S}{\Delta t_{st}}%
\int_{t_{1}}^{t_{2}}\;_{\pm }\bar{\psi}_{\mathbf{p}^{\prime }}(t)\mathbf{%
\gamma }\boldsymbol{\epsilon }_{\mathbf{K}\vartheta }\;_{\pm }\psi _{\mathbf{%
p}}(t)e^{i\omega t}dt\ ,  \label{e8}
\end{eqnarray}%
where $\delta _{\mathbf{p}^{\prime },\pm \mathbf{p}-\hbar \mathbf{k}}$ is
the Kronecker symbol, the spinor $_{\pm }\psi _{\mathbf{p}}(t)$ is given by
Eq.~(\ref{e6}), and it is taken into account that the contribution to the
integral over times $t\in \mathrm{I}\cup \mathrm{II}$ is zero due to the
absence of the electric field. Squaring the amplitudes (\ref{e8}) and
summing over the momenta $\mathbf{p}^{\prime }$, we represent the
probability densities (\ref{e7a3}) and (\ref{e7b}) as\textbf{\ }%
\begin{eqnarray}
&&\frac{d\mathcal{P}\left( \mathbf{p};\left. \mathbf{K,}\vartheta
\right\vert 0\right) }{d\omega d\Omega }=\frac{\alpha }{\varepsilon }\left(
\frac{v_{F}}{c}\right) ^{2}\frac{\omega \Delta t_{st}^{2}}{\left( 2\pi
\right) ^{2}}\left. \left\vert M_{\mathbf{p}^{\prime }\mathbf{p}%
}^{0}\right\vert ^{2}\right\vert _{\mathbf{p}^{\prime }=\mathbf{p}-\hbar
\mathbf{k}}\;,  \label{e9a} \\
&&\frac{d\mathcal{P}\left( \left. \mathbf{K},\vartheta \right\vert \overset{%
\pm }{\mathbf{p}}\right) }{d\omega d\Omega }=\frac{\alpha }{\varepsilon }%
\left( \frac{v_{F}}{c}\right) ^{2}\frac{\omega \Delta t_{st}^{2}}{\left(
2\pi \right) ^{2}}\left. \left\vert M_{\mathbf{p}^{\prime }\mathbf{p}}^{\pm
}\right\vert ^{2}\right\vert _{\mathbf{p}^{\prime }=\mathbf{p}-\hbar \mathbf{%
k}}\;.  \label{e9b}
\end{eqnarray}

Using the explicit representations (\ref{gamma}) and (\ref{e6}) as well as
the substitutions%
\begin{eqnarray}
&&u=\sqrt{\frac{v_{F}}{eE\hbar }}\left[ eEt-\frac{1}{2}\left(
p_{x}+p_{x}^{\prime }\right) \right] ,\quad u\left( t_{1,2}\right) =\left.
u\right\vert _{t=t_{1,2}},  \notag \\
&&u_{x}=\sqrt{\frac{v_{F}}{eE\hbar }}\left( p_{x}^{\prime }-p_{x}\right)
,\quad u_{\pm }=u\pm u_{x}/2,\;u_{0}=\Delta t_{st}\omega \ ,  \label{e10}
\end{eqnarray}%
we obtain%
\begin{eqnarray}
&&M_{\mathbf{p}^{\prime }\mathbf{p}}^{0}=-\exp \left( i\omega \frac{%
p_{x}+p_{x}^{\prime }}{2eE}\right) \exp \left[ -\frac{\pi \left( \lambda
+\lambda ^{\prime }\right) }{8}\right]  \notag \\
&&\times \left\{ i\chi _{\vartheta }^{0,1}\tilde{Y}_{00}+\left( 2eE\hbar
\right) ^{-1}v_{F}(mv_{F}+i\zeta p_{y}^{\prime })(mv_{F}+i\zeta p_{y})\chi
_{\vartheta }^{1,0}\tilde{Y}_{11}\right.  \notag \\
&&+\left. e^{-i\pi /4}\sqrt{\frac{v_{F}}{2eE\hbar }}\left[ -(mv_{F}+i\zeta
p_{y}^{\prime })\chi _{\vartheta }^{1,1}\tilde{Y}_{10}+(mv_{F}+i\zeta
p_{y})\chi _{\vartheta }^{0,0}\tilde{Y}_{01}\right] \right\} ,  \notag \\
&&M_{\mathbf{p}^{\prime }\mathbf{p}}^{+}=-\exp \left( i\omega \frac{%
p_{x}+p_{x}^{\prime }}{2eE}\right) \exp \left[ -\frac{\pi \left( \lambda
+\lambda ^{\prime }\right) }{8}\right]  \notag \\
&&\times \left\{ \chi _{\vartheta }^{0,0}Y_{00}+\left( 2eE\hbar \right)
^{-1}v_{F}(mv_{F}+i\zeta p_{y}^{\prime })(mv_{F}-i\zeta p_{y})\chi
_{\vartheta }^{1,1}Y_{11}\right.  \notag \\
&&+\left. \sqrt{\frac{v_{F}}{2eE\hbar }}e^{i\pi /4}\left[ (mv_{F}+i\zeta
p_{y}^{\prime })\chi _{\vartheta }^{1,0}Y_{10}-i(mv_{F}-i\zeta p_{y})\chi
_{\vartheta }^{0,1}Y_{01}\right] \right\} ,  \label{e11}
\end{eqnarray}%
where
\begin{eqnarray}
Y_{j^{\prime }j}\left( t_{2},t_{1}\right)
&=&\int_{u(t_{1})}^{u(t_{2})}D_{-\nu ^{\prime }-j^{\prime
}}[-(1+i)u_{-}]D_{\nu -j}[-(1-i)u_{+}]e^{iu_{0}u}du,  \label{e12a} \\
\tilde{Y}_{j^{\prime }j}\left( t_{2},t_{1}\right)
&=&\int_{u(t_{1})}^{u(t_{2})}D_{-\nu ^{\prime }-j^{\prime
}}[-(1+i)u_{-}]D_{-\nu -j}[-(1+i)u_{+}]e^{iu_{0}u}du,  \label{e12b}
\end{eqnarray}%
and%
\begin{equation*}
\chi _{\vartheta }^{\left( 1-s^{\prime }\right) /2,\left( 1-s\right)
/2}=U_{s^{\prime }}^{\dagger }\gamma ^{0}\overrightarrow{\gamma }\cdot \vec{%
\epsilon}_{\mathbf{K}\vartheta }U_{s},\;\nu ^{\prime }=\frac{i\lambda
^{\prime }}{2}\,,\quad \lambda ^{\prime }=\left. \lambda \right\vert
_{p_{y}\rightarrow p_{y}^{\prime }}\;.
\end{equation*}%
$\,$\ One can check that the probability density for an electron and hole in
Eq. (\ref{e9b}) are easily related by the replacement of $\mathbf{p}^{\prime
}\rightleftarrows \mathbf{p}$,%
\begin{equation}
M_{\mathbf{p}^{\prime }\mathbf{p}}^{-}=M_{\mathbf{pp}^{\prime }}^{+}\ .
\label{e13}
\end{equation}

To evaluate the angular matrix element $\chi _{\vartheta }^{j^{\prime },j}$,
we adopt the convention used in Ref. \cite{mecklenburg} and define an
orthonormal triple%
\begin{eqnarray}
&&\mathbf{K/}K=(\sin \theta \cos \phi ,\,\sin \theta \sin \phi ,\,\cos
\theta )\,,  \notag \\
&&\boldsymbol{\epsilon }_{\mathbf{K}1}=\mathbf{e}_{z}\times \mathbf{K/}%
\left\vert \mathbf{e}_{z}\times \mathbf{K}\right\vert ,\quad \boldsymbol{%
\epsilon }_{\mathbf{K}2}=\mathbf{K\times }\boldsymbol{\epsilon }_{\mathbf{K}%
1}/\left\vert \mathbf{K\times }\boldsymbol{\epsilon }_{\mathbf{K}%
1}\right\vert .  \label{e14}
\end{eqnarray}%
Then
\begin{eqnarray}
\boldsymbol{\epsilon }_{\mathbf{K}1} &=&(-\sin \phi ,\,\cos \phi ,\,0)\,,
\notag \\
\boldsymbol{\epsilon }_{\mathbf{K}2} &=&(-\cos \theta \cos \phi ,\,-\cos
\theta \sin \phi ,\,\sin \theta )\,  \label{e14b}
\end{eqnarray}%
for $\mathbf{K}$ in the upper spatial region, $k_{z}\geq 0$. Thereby, we
obtain:
\begin{align}
\chi _{1}^{1,1}& =-\chi _{1}^{0,0}=\sin \phi ,\ \chi _{1}^{0,1}=-\chi
_{1}^{1,0}=i\zeta \cos \phi \ ;  \notag \\
\chi _{2}^{1,1}& =-\chi _{2}^{0,0}=\cos \theta \cos \phi ,\ \chi
_{2}^{0,1}=-\chi _{2}^{1,0}=-i\zeta \cos \theta \sin \phi \ .  \label{e15}
\end{align}

For the momenta $p_{x}$ and $p_{x}^{\prime }$ satisfying condition (\ref{e3}%
) and for finite $u_{0}\ll \min \left( u(t_{1}),\left\vert
u(t_{2})\right\vert \right) $, it is possible to consider limits $%
T\rightarrow \infty $ in integrals (\ref{e12a}) and (\ref{e12b}). We denote
the corresponding limits as:%
\begin{equation}
Y_{j^{\prime }j}=\left. Y_{j^{\prime }j}\left( t_{2},t_{1}\right)
\right\vert _{T\rightarrow \infty },\;\tilde{Y}_{j^{\prime }j}=\left. \tilde{%
Y}_{j^{\prime }j}\left( t_{2},t_{1}\right) \right\vert _{T\rightarrow \infty
}\ .  \label{e16}
\end{equation}

These limits can be simplified using the hyperbolic coordinates $\rho $ and $%
\varphi $,%
\begin{equation}
\rho =\sqrt{u_{0}^{2}-u_{x}^{2}},\ \tanh \varphi =\frac{u_{x}}{u_{0}}\;%
\mathrm{if}\;u_{0}^{2}-u_{x}^{2}>0\ .  \label{e17}
\end{equation}%
We note that in both cases of the emission, we have $p_{x}^{\prime
}=p_{x}-\hbar k_{x}$. Therefore, in any frequency range the ratio $%
\left\vert u_{x}\right\vert /u_{0}$\ is very small,
\begin{equation}
\frac{\left\vert u_{x}\right\vert }{u_{0}}=\frac{\left\vert k_{x}\right\vert
}{K}\frac{v_{F}}{c}\leq \frac{v_{F}}{c},  \label{lim5}
\end{equation}%
and the condition $u_{0}^{2}-u_{x}^{2}>0$\ is fulfilled.{\large \ }This
feature of photon emission is due to the fact that the Fermi velocity $v_{F}$
in graphene is much smaller than the speed of light $c$.

The $\varphi $ dependence of integrals (\ref{e16}) can be factorized with
the help of Eq.~(\ref{a6}) (see Appendix \ref{App1}) and takes the form:%
{\large \ } \textrm{\ }
\begin{eqnarray}
&&Y_{j^{\prime }j}=\exp \left[ \left( i\frac{\lambda ^{\prime }-\lambda }{2}%
+j^{\prime }+j-1\right) \varphi \right] \mathcal{J}_{j^{\prime },j}(\rho ),
\notag \\
&&\mathcal{J}_{j^{\prime },j}(\rho )=\int_{-\infty }^{\infty }D_{-\nu
^{\prime }-j^{\prime }}[-(1+i)u]D_{\nu -j}[-(1-i)u]e^{i\rho u}du;
\label{e18a} \\
&&\tilde{Y}_{j^{\prime }j}=\exp \left[ \left( i\frac{\lambda ^{\prime
}-\lambda }{2}+j^{\prime }-j\right) \varphi \right] \mathcal{\tilde{J}}%
_{j^{\prime },j}(\rho ),  \notag \\
&&\mathcal{\tilde{J}}_{j^{\prime },j}(\rho )=\int_{-\infty }^{\infty
}D_{-\nu ^{\prime }-j^{\prime }}[-(1+i)u]D_{-\nu -j}[-(1+i)u]e^{i\rho u}du.
\label{e18b}
\end{eqnarray}%
These integrals can be expressed via the confluent hypergeometric function $%
\Psi $ as
\begin{eqnarray}
&&\mathcal{J}_{j^{\prime },j}(\rho )=\left( -1\right) ^{j}\sqrt{\frac{2}{\pi
}}\Gamma \left( \nu -j+1\right) e^{i\pi \left( \nu ^{\prime }+j^{\prime
}-1\right) /2}\sinh \frac{\pi \lambda }{2}I_{j^{\prime },1-j}(\rho ),\
\label{e21b} \\
&&\mathcal{\tilde{J}}_{j^{\prime },j}(\rho )=e^{i\pi \left( \nu +\nu
^{\prime }+j+j^{\prime }\right) /2}I_{j^{\prime },j}(\rho ),\ \
\label{e21a} \\
&&I_{j^{\prime },j}(\rho )=\sqrt{\pi }\exp \left[ \left( \ln \frac{\rho }{%
\sqrt{2}}-\frac{i\pi }{4}\right) \left( \nu -\nu ^{\prime }+j-j^{\prime
}\right) +i\frac{\rho ^{2}}{2}-\frac{i\pi }{4}\right]  \notag \\
&&\times \Psi \left( \nu +j,1+\nu -\nu ^{\prime }+j-j^{\prime };-i\frac{\rho
^{2}}{2}\right) ,  \label{e21c}
\end{eqnarray}%
where $\Gamma $ is the gamma-function (see Appendix \ref{App1} for details).

\subsection{Natural limits of parameters\label{SS3.3}}

In the case under consideration, one should take into account the existence
of natural limits of physical parameters characterizing both the charged
particles themselves and their radiation.

Let us consider the domain of the applicability of the perturbation theory
with respect of the photon emission in the case under consideration. In the $%
T$-constant electric field there is the natural range of the very low
frequency of emission, $\omega \lesssim \omega ^{\mathrm{IR}}=2\pi T^{-1}$.%
{\large \ }In this range the perturbation theory works if the total number
of photons is small enough.{\large \ }Otherwise, the radiation must be
treated in the mean field approximation. For our purposes, it is enough to
restrict the applicability of the perturbation theory with respect of the
photon emission by the condition $\omega >\omega ^{\mathrm{IR}}$, which is
convenient to represent as:%
\begin{equation}
u_{0}>u_{0}^{\mathrm{IR}},\;u_{0}^{\mathrm{IR}}=2\pi \Delta t_{st}T^{-1}.
\label{IR1}
\end{equation}

It should be recalled that, in a number of cases, the need to cut off from
below the region of the radiation frequencies is often encountered in QED
problems.{\large \ }This makes it possible to deal with divergences (the
well known infrared catastrophe) whose nature is associated with the
impossibility of separating a charged particle from its radiation field; see
Appendix \ref{App2} for details. It is known that such soft photons carry
away only a negligibly small part of the energy of emission, so that the
corresponding back reaction is also negligible.{\large \ }In the case under
consideration an estimation of the corresponding cut off parameter shows
that its value is much less than the quantity $u_{0}^{\mathrm{IR}}$, which
means, in turn, that the domain of the applicability of the perturbation
theory is bigger than the one that follows from the inequality (\ref{IR1}).%
{\large \ }Therefore, condition (\ref{IR1}) provides the possibility of
applicability of the perturbation theory in the problem we are considering.%
{\large \ }Thus, we believe that results obtained in section (\ref{SS3.2})
may be considered credible in all the frequency range (\ref{IR1}).

We note now that condition (\ref{IR1}){\large \ }is not a significant
limitation when applying our approach to a wide class of similar physical
problems.{\large \ }Indeed, as has been shown in Ref. \cite{GavGitY12}
nonlinear and linear $I-V$\ graphs\ experimentally observed in low and
high-mobility graphene samples \cite{vandecasteele}, can be explained in the
framework of strong-field QED$_{3,2}$ in the mean field approximation,
taking into account the backreaction of the mean current of created carriers
to the applied electric field which is set by a constant voltage.{\large \ }%
In addition, it has been found that the radiation of a time-dependent mean
current, forming the backreaction to the electric field on the graphene
plane,\ is emitted to the three-dimensional space in the form of very low
frequency, $\omega \lesssim \omega ^{\mathrm{IR}}$, linearly polarized plane
electromagnetic waves.{\large \ }It can be seen that the backreaction of the
mean current can be neglected on the big time intervals of the time scale $%
\Delta t_{st}$ order,{\large \ }which is equivalent to the assumption that
the electric field is constant.{\large \ }Thus, the backreaction does not
effect the emission of high-frequency photons, because the corresponding
formation interval $\Delta t$\ is of the order $\Delta t_{st}$\ (the latter
will be demonstrated below).

Maximum possible values of particle momenta in the $T$-constant field were
determined by the equation (\ref{e3}). The finite dimensions of graphene
samples do not allow us to consider the spectrum of small momenta to be
continuous. However, the dependence on the longitudinal impulses $p_{x}$ in
the expressions (\ref{e11}), (\ref{e12a}), and (\ref{e12b}) is such that the
discreteness of these momenta can simply be ignored. Nevertheless, for small
lateral width $L_{y}$, only those momenta $p_{y}$ that are not very small,
namely, satisfy the condition%
\begin{equation*}
\left\vert p_{y}\right\vert \gg \Delta p_{y},\;\Delta p_{y}=\frac{2\pi \hbar
}{L_{y}}.
\end{equation*}%
In turn, this this limits from below the admissible values of the
dimensionless parameter $\lambda $,%
\begin{equation}
\sqrt{\lambda }\gg \sqrt{\lambda _{\min }},\;\sqrt{\lambda _{\min }}=\sqrt{%
\frac{v_{F}\hbar }{eE}}\frac{2\pi }{L_{y}}.  \label{lim1}
\end{equation}

In the absence of nanoribbons, we may assume that $L_{y}\sim 1\mathrm{\mu m}$%
, then%
\begin{equation}
\lambda _{\min }\sim \frac{2.7}{a}\times 10^{-2},  \label{lim2}
\end{equation}%
where the range of allowable values of dimensionless parameter $a$ is given
by Eq.~(\ref{g9}). In the case of nanoribbons one has to take into account
that typical width of a nanoribbon $l_{y}$ is $l_{y}\sim 1\mathrm{nm}$ and
the parameter $\lambda $ is quantized,%
\begin{equation}
\sqrt{\lambda _{n}}=\sqrt{\frac{v_{F}\hbar }{eE}}\frac{2\pi }{l_{y}}%
n,\;n=0,1,2,\ldots \ .  \label{lim2b}
\end{equation}%
Thus, for the smallest nonzero value of $\lambda _{1}$ we have: $\sqrt{%
\lambda _{1}}\sim 27/a$.

The large time scale is:%
\begin{equation}
\Delta t_{st}\approx \left( a\right) ^{-1/2}2.6\times 10^{-14}\mathrm{s\ }.
\label{t-scale}
\end{equation}%
Characteristic frequency%
\begin{equation}
\omega _{sc}=\Delta t_{st}^{-1}\approx \sqrt{a}\times 0.39\times 10^{14}%
\mathrm{s}^{-1},  \label{f-scale}
\end{equation}%
provides a value of one for dimensionless parameter $u_{0}$ and, therefore,
specifies a frequency scale against which high-frequency and low-frequency
emissions regions can be defined. Note that it depends on electric field $E$
value. Characteristic wavelength scale is:%
\begin{equation}
l_{sc}=\frac{2\pi c}{\omega _{sc}}\approx \frac{48}{\sqrt{a}}\times 10^{-6}%
\mathrm{m}  \label{l-scale}
\end{equation}%
For example, in the case of the typical voltage $\sim 1\,\mathrm{V}$ ($a\sim
1$) the corresponding high-frequency range is a mid-wavelength infrared. We
stress that terahertz-field induced spontaneous optical emission in the
range of 340--600 nm was observed from a monolayer graphene on a glass
substrate \cite{emis-exp17,emis-exp21}.

The differential probabilities (\ref{e7a2}) and (\ref{e7a1}) can be
integrated over $\mathbf{K}$ only between such limits that leave the
integral probability much smaller than unity. Let us demonstrate that for
the integration over $\omega $ there is a natural cutoff from above. Let us
consider the high frequency case,%
\begin{equation}
u_{0}\gtrsim \tau _{\gamma },\;\;\tau _{\gamma }\gg 1,  \label{lim3}
\end{equation}%
where $\tau _{\gamma }$ is an arbitrary given number. For probability
densities of the photon emission with $\mathbf{K}$, given by Eqs. (\ref{e9a}%
) and (\ref{e9b}), an important role is played by definite time intervals (%
\ref{e12a}) and (\ref{e12b}). These are intervals where main contributions
to the integrals are formed. On the same intervals the main contributions
are formed to probability densities of the photon absorption. We can find
this intervals using the saddle-point method.

Let us consider integral (\ref{e12b}). Under condition (\ref{lim3}), the
mentioned saddle-point is situated in the range where absolute values of
arguments of both WPCF's involved in integral (\ref{e12b}) are big,
\begin{equation}
\left\vert u+u_{x}/2\right\vert \gg \max \left\{ 1,\lambda \right\} \;%
\mathrm{and}\;\left\vert u-u_{x}/2\right\vert \gg \max \left\{ 1,\lambda
\right\} .  \label{lim3b}
\end{equation}%
In this case, if $u\pm u_{x}/2<0$, one uses the following asymptotic
expansion:

\begin{equation}
D_{p}\left( z\right) =e^{-z^{2}/4}z^{p}\left[ 1+O\left( \left\vert
z\right\vert ^{-2}\right) \right] \;\;\mathrm{if}\;\left\vert \arg
z\right\vert <\frac{3\pi }{4}.  \label{asy_exp}
\end{equation}%
If $u\pm u_{x}/2>0,$ applying Eq. (\ref{asy_exp}), one uses a relation
between WPCF's (see (2.8.2.(7)) in Ref. \cite{HTF2}),%
\begin{equation}
D_{p}\left( z\right) =e^{-i\pi p}D_{p}\left( -z\right) +\frac{\sqrt{2\pi }}{%
\Gamma \left( -p\right) }e^{-i\pi \left( p+1\right) /2}D_{-p-1}\left(
iz\right) .  \label{PCFrel}
\end{equation}%
Thus one finds that the saddle-point is $u=u_{0}/2$. Since $u_{0}$ is
positive, the saddle-point can be situated only in the range $u\pm u_{x}/2>0$%
. Following the same way one finds that the saddle-point of the kernel in
integral (\ref{e12a}) is $u=u_{0}/2$ and is also situated in the range $u\pm
u_{x}/2>0$.

Using substitutions (\ref{e10}) one can see that the saddle-point equation
represents a conservation law of the kinetic energy,
\begin{equation}
v_{F}\left[ 2eEt-\left( p_{x}+p_{x}^{\prime }\right) \right] =\hbar \omega .
\label{en_cons}
\end{equation}%
In the neighborhood of the saddle-point the corresponding kernels have
Gaussian forms with maxima at the time instant%
\begin{equation}
t_{c}=\frac{1}{2}\left( \Delta t_{st}^{2}\omega +\frac{p_{x}+p_{x}^{\prime }%
}{eE}\right)  \label{tc}
\end{equation}%
and with the standard deviation
\begin{equation}
\Delta t_{sd}=\Delta t_{st}/\sqrt{2}.  \label{sd}
\end{equation}%
The time $t_{c}$ corresponds to the position of the center of the formation
interval $\Delta t$ for given $\omega $, $p_{x}$, and $p_{x}^{\prime }$. The
width of the formation interval $\Delta t$ must be large enough to
accommodate the points $u+u_{x}/2$ and $u-u_{x}/2$. In addition, the
formation interval must overlap the interval $\Delta t_{sd}$, $\Delta
t_{sd}<\Delta t$. It is natural to assume that $\Delta t\sim \Delta t_{st}$.
It implies the following condition:%
\begin{equation}
\left\vert u_{x}\right\vert <1\ .  \label{lim4}
\end{equation}%
With account taken of the relation $p_{x}^{\prime }=p_{x}-\hbar k_{x}$\ one
can see that inequality (\ref{lim4}) implies:%
\begin{equation}
v_{F}\Delta t_{st}\left\vert k_{x}\right\vert =\frac{\left\vert
k_{x}\right\vert }{K}\frac{v_{F}}{c}u_{0}<1.  \label{lim5b}
\end{equation}%
{\large \ }Thus, in the case of high frequencies, the width of the formation
interval does not depend on the frequency $\omega $ and on the momentum of
the particle and is determined entirely by the electric field $E$. Thus,
the\ variation\ of\ the\ external electric\ field\ acting\ on\ the\
particle\ within\ the\ formation\ length\ can\ be\ neglected,\ which allows
us to use the\ locally\ constant\ field\ approximation.{\large \ }By the
same reason, the obtained results can be easily extended to the study of the
emission in any slowly varying field configuration. Assuming that the
electric field $E$ decreases quickly enough beyond the formation interval
(for example, as a result of the backreaction of created pairs, see details
in Ref. \cite{GavGitY12}), the upper limitation (\ref{g9}) to the intensity
of the constant electric field can be significantly weakened. This means,
for example, that the above considerations may be extended to terahertz
pulses of intensity from 100 to 250 $kV/cm$, which are used in the existing
experiments \cite{emis-exp17,emis-exp21}.

The saddle-point of the kernel involved in integral (\ref{e12a}) is located
in the range $u\pm u_{x}/2>0$, that is, longitudinal kinetic momenta of an
emitting electron are negative, $P_{x}\left( t\right) =p_{x}-eEt<0$ and $%
P_{x}^{\prime }\left( t\right) =p_{x}^{\prime }-eEt<0$ (longitudinal kinetic
momenta of a hole are positive, $-P_{x}\left( t\right) $ and $-P_{x}^{\prime
}\left( t\right) $). According to Eq. (\ref{lim5}), $P_{x}$ and $%
P_{x}^{\prime }$ differ little in magnitude, so we can neglect the
contribution from the small longitudinal component $k_{x}$ in Eq. (\ref%
{en_cons}), which gives: $2v_{F}\left\vert P_{x}\left( t\right) \right\vert
\approx \hbar \omega $. Note that in the $T$-constant field, the range of $%
p_{x}$, given by Eq. (\ref{e3}), implies that the initial kinetic momenta of
an electron under consideration are always positive, $P_{x}\left(
t_{1}\right) >0$ (initial kinetic momenta of a hole are negative). In the
case of the photon emission which accompanies the pair production from the
initial vacuum, the saddle-point of the kernel involved in integral (\ref%
{e12b}) is located in the same range $u\pm u_{x}/2>0$, that is, the
longitudinal kinetic momentum of the electron of a pair is negative, $%
P_{x}^{\prime }\left( t\right) <0$, while the longitudinal kinetic momentum
of the hole of a pair is positive, $P_{x}^{\left( h\right) }\left( t\right)
=-P_{x}\left( t\right) >0$. The conservation law for the kinetic energy at
the saddle-point, given by Eq. (\ref{en_cons}), can be written in terms of
these momenta as $v_{F}\left[ P_{x}^{\left( h\right) }\left( t\right)
+\left\vert P_{x}^{\prime }\left( t\right) \right\vert \right] =\hbar \omega
$. It follows from Eq. (\ref{tc}) that%
\begin{equation}
t_{c}\approx \frac{1}{2}\Delta t_{st}^{2}\omega +\frac{p_{x}}{eE}\ .
\label{lim6}
\end{equation}%
It means that for the photon emission of a given frequency\emph{\ }$\omega $%
\emph{\ }dependence of the effect on\emph{\ }$p_{x}$\emph{\ }comes down to
just shifting of the center of the formation interval.\emph{\ }On the other
hand, for a given momentum $p_{x}$ photons with high frequencies are formed
later.

It follows from Eq. (\ref{lim6}) that for any given $p_{x}$ the high
frequency emission, $\omega /\omega _{sc}\gtrsim \tau _{\gamma }$, starts
when the longitudinal kinetic momentum $P_{x}\left( t\right) $ reaching its
threshold value at $t_{c}\sim t_{0}$ according to condition (\ref{lim3}),%
\begin{equation}
\frac{2\left\vert P_{x}\left( t_{0}\right) \right\vert }{eE\Delta t_{st}}%
\approx \tau _{\gamma }.  \label{lim7}
\end{equation}%
The minimal frequency where the region of high frequencies starts is:
\begin{equation}
\omega _{\min }=\frac{2\left\vert P_{x}\left( t_{0}\right) \right\vert }{%
eE\Delta t_{st}}\omega _{sc}\approx \tau _{\gamma }\omega _{sc}\ .
\label{e34}
\end{equation}

The smallest possible value of the moment $t_{0}$ , at which Eq. (\ref{e34})
is satisfied, is achieved at the smallest possible momentum value $p_{x}$
from the finite range (\ref{e3}). Taking it into account, we find%
\begin{equation}
t_{0}-t_{1}\sim \left( \tau _{\gamma }/2+\tau \right) \Delta t_{st}.
\label{e36}
\end{equation}

The frequency $\omega $ grows from the minimum value $\omega _{\min }$ as
long as the electric field is acting and reaches the maximum possible for a
given $p_{x}$ frequency $\omega _{2}$ at the time instant $t_{c}\sim t_{2}$,
when the electric field switches off. Photon with such a frequency is
emitted during the formation interval preceding the moment $t_{2}$ of
switching off the electric field. It follows from Eq. (\ref{lim6}) that%
\begin{equation}
\omega _{2}\approx \frac{2\left\vert P_{x}\left( t_{2}\right) \right\vert }{%
eE\Delta t_{st}}\omega _{sc}.  \label{e32}
\end{equation}%
Absolute maximum among all possible frequencies $\omega _{2}$ with different
momenta $p_{x}$ satisfying Eq. (\ref{e3}) is:%
\begin{equation}
\omega _{\mathrm{\max }}\approx 2\Delta t_{st}^{-2}\left[ t_{2}+\max \left(
-p_{x}/eE\right) \right] \approx 2\left( T/\Delta t_{st}\right) \omega
_{sc}\ .  \label{e33}
\end{equation}%
A frequency range between $\omega _{\mathrm{\max }}$ and $\omega _{\min }$
does exists if
\begin{equation}
\frac{\omega _{\mathrm{\max }}}{\omega _{\min }}\approx \frac{2T}{\Delta
t_{st}\tau _{\gamma }}>1,  \label{e33a}
\end{equation}%
which means that the field duration time $T$ satisfying Eqs. (\ref%
{time-condition}) and (\ref{g8}) is sufficiently large.

In particular, it follows from the estimation (\ref{e33}) that dimensionless
parameter $u_{0}$ is restricted from above,%
\begin{equation}
u_{0}<\omega _{2}\Delta t_{st}<u_{0}^{\max },\;\;u_{0}^{\max }=\omega _{%
\mathrm{\max }}/\omega _{sc}\approx 2T/\Delta t_{st}\ .  \label{e33b}
\end{equation}%
Note, that if%
\begin{equation}
\frac{v_{F}}{c}u_{0}^{\max }<1,  \label{lim5c}
\end{equation}%
inequality (\ref{lim5b}) always holds true. Besides, the lower bound of the
range of the frequency, given by Eq. (\ref{IR1}), is also defined by the
quantity $u_{0}^{\max }${\large , }%
\begin{equation}
u_{0}^{\mathrm{IR}}=4\pi /u_{0}^{\max }.  \label{IR2}
\end{equation}%
Choosing $T=T_{bal}\sim 10^{-12}\mathrm{s}$ and taking into account the
estimation for the quantity $\Delta t_{st}$ given by Eq.(\ref{t-scale}), we
find%
\begin{equation}
u_{0}^{\max }\approx 78\sqrt{a},\;\omega _{\mathrm{\max }}\approx 3.0\times
10^{15}a\;\mathrm{s}^{-1}\ .  \label{e33d}
\end{equation}

We believe that at $\tau _{\gamma }\sim 3$ one can confident enough to
identify the range of high frequencies. It follows from estimation (\ref%
{e33d}) and from restrictions on the parameter $a$ given by Eq.~(\ref{g9})
that the high frequency range $u_{0}\gtrsim \tau _{\gamma }$ definitely
exists for the fields under consideration.

\section{High frequency approximation \label{S4}}

In section \ref{SS3.2}, we have obtained characteristics of the one-photon
emission probabilities that are valid in range (\ref{IR1}).{\large \ }In the
general case, angular and polarization distributions of the emitted photons
have quite complicated form.{\large \ }Nevertheless,{\large \ }their
analysis is greatly simplified in the range of high frequencies (we recall
that this range is defined by relation (\ref{lim3})).{\large \ }One can see
that namely the emission in this range makes the main contribution to the
one-photon emission considered by us.{\large \ }This is explained by the
fact that in this case, the width of the formation interval is small $\Delta
t\sim \Delta t_{st}${\large \ }and does not depend on the frequency $\omega $%
\ and on the particles momenta and is determined entirely by the electric
field $E$.{\large \ }Thus, the obtained results can be extended to the study
of the emission in any slowly varying field configuration.{\large \ }For
this reason the emission of high-frequency photons which accompanies the
electronic quantum transport in the graphene is more realistic for possible
experimental observations. In which follows, we assume that the mass gap in
the graphene is absent, $m=0$, and we neglect the small terms depending on $%
\varphi $. At high frequencies,%
\begin{equation}
\rho \approx u_{0}>\tau _{\gamma },  \label{e23}
\end{equation}%
and using an asymptotic behavior of the function $\Psi $ given by Eq.
(6.13.1.(1)) in Ref. \cite{HTF1}, we find:%
\begin{equation*}
\Psi \left( \nu +j,1+\nu -\nu ^{\prime }+j-j^{\prime };-i\frac{\rho ^{2}}{2}%
\right) =\left( -i\frac{\rho ^{2}}{2}\right) ^{-\nu -j}\left[ 1+O\left( \rho
^{-2\left( j+1\right) }\right) \right] \ .
\end{equation*}%
Whence it follows that

\begin{equation}
I_{j^{\prime }j}\left( \rho \right) \approx \sqrt{\pi }\left( \frac{\rho }{%
\sqrt{2}}\right) ^{-\nu ^{\prime }-\nu -j^{\prime }-j}\exp \left\{ \frac{i}{4%
}\left[ -\pi +\rho ^{2}+\pi \left( \nu ^{\prime }+\nu +j^{\prime }+j\right) %
\right] \right\}  \label{e24}
\end{equation}

We see that the leading contribution to the amplitude $M_{\mathbf{p}^{\prime
}\mathbf{p}}^{+}$ given by Eq. (\ref{e11}) is due to the term with $Y_{01}$.
Using representations (\ref{e18a}), (\ref{e21a}), and (\ref{e24}), we find:
\begin{eqnarray}
&&\left. \left\vert M_{\mathbf{p}^{\prime }\mathbf{p}}^{+}\right\vert
^{2}\right\vert _{\mathbf{p}^{\prime }=\mathbf{p}-\hbar \mathbf{k}}\;\approx
f\left( \lambda ,\lambda ^{\prime }\right) \left\vert \chi _{\vartheta
}^{0,1}\right\vert ^{2},\;  \notag \\
&&f\left( \lambda ,\lambda ^{\prime }\right) =2\pi \sinh \frac{\pi \lambda }{%
2}\exp \left[ -\frac{\pi }{4}\left( 5\lambda +7\lambda ^{\prime }\right) %
\right] \ ,  \label{e25}
\end{eqnarray}%
where $\chi _{\vartheta }^{0,1}$ is given by Eq. (\ref{e15}) and%
\begin{equation}
\lambda ^{\prime }=\left( v_{F}\Delta t_{st}\right) ^{2}\left\vert
k_{y}-p_{y}/\hbar \right\vert ^{2}.  \label{e26}
\end{equation}%
Thus, we find that the asymptotic behavior of the probability of the one
photon emission with a given polarization $\vartheta $ from a
single-electron (hole) state per unit frequency and solid angle is:%
\begin{equation}
\frac{d\mathcal{P}\left( \left. \mathbf{K},\vartheta \right\vert \overset{%
\pm }{\mathbf{p}}\right) }{d\omega d\Omega }=\frac{\alpha }{\varepsilon }%
\left( \frac{v_{F}}{c}\right) ^{2}\frac{\omega \Delta t_{st}^{2}}{\left(
2\pi \right) ^{2}}\left. \left\vert M_{\mathbf{p}^{\prime }\mathbf{p}}^{\pm
}\right\vert ^{2}\right\vert _{\mathbf{p}^{\prime }=\mathbf{p}-\hbar \mathbf{%
k}}\;,  \label{e27}
\end{equation}%
where $\left\vert M_{\mathbf{p}^{\prime }\mathbf{p}}^{+}\right\vert ^{2}$ is
given by Eq. (\ref{e25}) and $\left\vert M_{\mathbf{p}^{\prime }\mathbf{p}%
}^{-}\right\vert ^{2}=$ $\left\vert M_{\mathbf{pp}^{\prime }}^{+}\right\vert
^{2}$. Summing probabilities (\ref{e27}) over the polarizations, we obtain
the probability of unpolarized emission from a single-electron (hole) state
per unit frequency and solid angle as:
\begin{eqnarray}
&&\frac{d\mathcal{P}\left( \left. \mathbf{K}\right\vert \overset{\pm }{%
\mathbf{p}}\right) }{d\omega d\Omega }=\frac{\alpha }{\varepsilon }\left(
\frac{v_{F}}{c}\right) ^{2}\frac{\omega \Delta t_{st}^{2}}{\left( 2\pi
\right) ^{2}}\left. \left\vert \tilde{M}_{\mathbf{p}^{\prime }\mathbf{p}%
}^{\pm }\right\vert ^{2}\right\vert _{\mathbf{p}^{\prime }=\mathbf{p}-\hbar
\mathbf{k}}\;,\;\left\vert \tilde{M}_{\mathbf{p}^{\prime }\mathbf{p}%
}^{-}\right\vert ^{2}=\left\vert \tilde{M}_{\mathbf{pp}^{\prime
}}^{+}\right\vert ^{2},  \notag \\
&&\left. \left\vert \tilde{M}_{\mathbf{p}^{\prime }\mathbf{p}%
}^{+}\right\vert ^{2}\right\vert _{\mathbf{p}^{\prime }=\mathbf{p}-\hbar
\mathbf{k}}\approx f\left( \lambda ,\lambda ^{\prime }\right) \left[ 1-\sin
^{2}\phi \left( 1-\cos ^{2}\theta \right) \ \right] \;.  \label{e28}
\end{eqnarray}

Probabilities (\ref{e27}) and (\ref{e28}) increase monotonically with
increasing the frequency $\omega $ and reach their maxima, given by Eqs. (%
\ref{e32}) and (\ref{e33}) respectively, as $\omega \rightarrow \omega
_{2}<\omega _{\mathrm{\max }}$ . One can find the probability of one-photon
emission from given distributions of electrons and holes of one kind per
unit frequency and solid angle as follows:%
\begin{equation}
\frac{d\mathcal{P}\left( \left. \mathbf{K},\vartheta \right\vert \overset{%
\pm }{\mathrm{in}}\right) }{d\omega d\Omega }=\frac{S}{\left( 2\pi \hbar
\right) ^{2}}\int \frac{d\mathcal{P}\left( \left. \mathbf{K},\vartheta
\right\vert \overset{\pm }{\mathbf{p}}\right) }{d\omega d\Omega }N_{\mathbf{p%
}}^{(\pm )}(\mathrm{in})dp_{x}dp_{y},  \label{e29}
\end{equation}%
where $N_{\mathbf{p}}^{(\pm )}(\mathrm{in})$ are some initial differential
mean numbers of electrons ($+$) and holes ($-$). If the numbers $N_{\mathbf{p%
}}^{(\pm )}(\mathrm{in})$ \ are the same for all the charge species, the
final probability is given by Eq. (\ref{e29}) multiplying it by the number $%
N_{f}$ of the species.

One can see that for given angles $\theta $\ and $\phi $ the function
defined by Eq. (\ref{e25}) has the Gaussian form as a function of the wave
number $k_{y}$, and besides $p_{y}/\hbar $ \ is the position of its maximum
and $\left( \sqrt{7\pi }v_{F}\Delta t_{st}/\sqrt{2}\right) ^{-1}$ \ is the
corresponding standard deviation. Note that this deviation increases with
the intensity of the electric field. We note that the emission from a
one-electron state depends essentially on the electron transversal momentum $%
p_{y}$. This emission takes place only if the latter momentum differs from
zero. For big $\lambda \gg 1$ the probability of the emission decreases
exponentially as $\lambda $ increases. The emission is maximum at $\tanh
\frac{\pi \lambda }{2}=\frac{2}{5}$, that is, the main contribution to it
comes from electrons with moderate magnitude $\lambda \sim 1/\pi .$ In this
case function (\ref{e25}) as the function of the wave number $k_{y}$ reaches
its maximum at $\lambda ^{\prime }=0$. It is convenient to introduce the
quantity $\omega _{y}=c\left\vert k_{y}\right\vert $, which represents a
corresponding contribution to the frequency $\omega $. It is obviously that $%
\omega _{y}<\omega $. Since $\omega <\omega _{\max }$, where $\omega _{\max
} $ is given by Eq. (\ref{e33}), we obtain the restriction $\omega
_{y}<\omega _{\max }$. The condition $\lambda ^{\prime }=0$ implies:
\begin{equation}
\frac{v_{F}}{c}\frac{\omega _{y}}{\omega _{sc}}=\sqrt{\lambda }\ ,
\label{e26b}
\end{equation}%
where $\omega _{sc}$ is given by Eq. (\ref{f-scale}). Assuming $\sqrt{%
\lambda }\sim 1$ one can satisfy condition (\ref{e26b}) only in the case
when $\omega _{\max }$ \ is big enough, such that
\begin{equation}
\frac{2Tv_{F}}{\Delta t_{st}c}\sim 1\text{.}  \label{e26c}
\end{equation}%
It is possible if the dimensionless parameter $T/\Delta t_{st}$ is big
enough as well. Otherwise the condition $\lambda ^{\prime }=0$ is
unreachable if $\sqrt{\lambda }\sim 1$ .

We see that the leading contribution to the amplitude $M_{\mathbf{p}^{\prime
}\mathbf{p}}^{0}$ given by Eq. (\ref{e11}) arrises from the term with $%
\tilde{Y}_{00}$. Using representations (\ref{e18b}), (\ref{e21a}), and (\ref%
{e24}), we find:%
\begin{equation}
\left. \left\vert M_{\mathbf{p}^{\prime }\mathbf{p}}^{0}\right\vert
^{2}\right\vert _{\mathbf{p}^{\prime }=-\mathbf{p}-\hbar \mathbf{k}%
}\;\approx \tilde{f}\left( \lambda ,\lambda ^{\prime }\right) \left\vert
\chi _{\vartheta }^{0,1}\right\vert ^{2},\;\tilde{f}\left( \lambda ,\lambda
^{\prime }\right) =\pi e^{-\pi \left( \lambda +\lambda ^{\prime }\right) },
\label{e30}
\end{equation}%
where $\chi _{\vartheta }^{0,1}$ is given by Eq. (\ref{e15}). We note that
the quantity $\tilde{f}\left( \lambda ,\lambda ^{\prime }\right) $ is
proportional to the product of differential mean numbers of electron and
hole of a pair created, respectively.{\large \ }Thus, we find that the
asymptotic behavior of the probability of the one-photon emission with a
given polarization $\vartheta $, which accompanies the production from the
initial vacuum state of pairs of charged species with a given momentum $%
\mathbf{p}$ per unit frequency and solid angle reads:%
\begin{equation}
\frac{d\mathcal{P}\left( \mathbf{p};\left. \mathbf{K,}\vartheta \right\vert
0\right) }{d\omega d\Omega }\approx \frac{\alpha }{\varepsilon }\left( \frac{%
v_{F}}{c}\right) ^{2}\frac{\omega \Delta t_{st}^{2}}{\left( 2\pi \right) ^{2}%
}\left. \left\vert M_{\mathbf{p}^{\prime }\mathbf{p}}^{0}\right\vert
^{2}\right\vert _{\mathbf{p}^{\prime }=\mathbf{p}-\hbar \mathbf{k}}\;,
\label{e31}
\end{equation}%
where $\left\vert M_{\mathbf{p}^{\prime }\mathbf{p}}^{0}\right\vert ^{2}$ is
given by Eq. (\ref{e30}). The total probability of the one-photon emission
with quantum numbers $\mathbf{K}$ and $\vartheta $, which accompanies the
pair production of all $N_{f}$ species from the initial vacuum state per
unit frequency and solid angle is presented by an integral over the finite
momentum range $D$, given by Eq. (\ref{e3}),
\begin{equation}
\frac{d\mathcal{P}\left( \mathbf{K,}\vartheta \right) }{d\omega d\Omega }=%
\frac{N_{f}S}{\left( 2\pi \hbar \right) ^{2}}\int_{D}\frac{d\mathcal{P}%
\left( \mathbf{p};\left. \mathbf{K,}\vartheta \right\vert 0\right) }{d\omega
d\Omega }dp_{x}dp_{y}\ .  \label{e32a}
\end{equation}%
Its asymptotic behavior has the form:%
\begin{eqnarray}
&&\frac{d\mathcal{P}\left( \mathbf{K,}\vartheta \right) }{d\omega d\Omega }%
\approx \mathcal{R}\left( \mathbf{K,}\vartheta \right) ST,\;\mathcal{R}%
\left( \mathbf{K,}\vartheta \right) =d\left( \omega ,\omega _{y}\right)
\left\vert \chi _{\vartheta }^{0,1}\right\vert ^{2},  \notag \\
&&d\left( \omega ,\omega _{y}\right) =\frac{\alpha N_{f}}{\varepsilon
2^{5/2}\pi }\frac{\omega }{\omega _{sc}l_{sc}^{2}}\exp \left[ -\frac{\pi }{2}%
\left( \frac{v_{F}\omega _{y}}{c\omega _{sc}}\right) ^{2}\right] ,
\label{e32b}
\end{eqnarray}%
where $l_{sc}$ is the characteristic wavelength scale given by Eq. (\ref%
{l-scale}). Note that probability (\ref{e32b}) is proportional to the total
number density of electron-hole pairs created, given by Eq. (\ref{n-cr}).
Summing the total probabilities (\ref{e32b}) over the polarizations, we
obtain the probability of unpolarized emission which accompanies the pair
production from the vacuum per unit frequency and solid angle:%
\begin{equation}
\frac{d\mathcal{P}\left( \mathbf{K}\right) }{d\omega d\Omega }\approx
\mathcal{R}\left( \mathbf{K}\right) ST,\;\mathcal{R}\left( \mathbf{K}\right)
=d\left( \omega ,\omega _{y}\right) \left[ 1-\sin ^{2}\phi \left( 1-\cos
^{2}\theta \right) \ \right] .  \label{e35}
\end{equation}%
The formula (\ref{e35}) was previously obtained in Ref. \cite{Yok14} for $%
\varepsilon =1$\ in the framework of many-body quantum mechanics, where the
interaction with external electric field is treated nonperturbatively.

Note that the frequency $\omega $ in Eqs. (\ref{e31}), (\ref{e32b}), and (%
\ref{e35}) is restricted from the above, $\omega <\omega _{\max }$, where $%
\omega _{\max }$ is given by Eq. (\ref{e33}). This implies the restriction $%
\omega _{y}<\omega <\omega _{\max }$ for $\omega _{y}$\ . Therefore,%
\begin{equation*}
\frac{v_{F}\omega _{y}}{c\omega _{sc}}<\frac{2Tv_{F}}{\Delta t_{st}c}\ .
\end{equation*}%
One can see that the argument of the exponential function in Eq. (\ref{e32b}%
) can significantly affect the value of the probability only under the
condition%
\begin{equation}
\frac{2Tv_{F}}{\Delta t_{st}c}\gtrsim 1.  \label{e37}
\end{equation}%
In this case one can see that this probability decreases exponentially if $%
\omega _{y}\rightarrow \omega $ and $\omega \rightarrow \omega _{\max }$,
that is, for frequencies close to the maximum $\omega _{\max }$, the
emission in the $y$-axis direction is suppressed. The probabilities (\ref%
{e32b}) and (\ref{e35}) increase monotonically with increasing frequency $%
\omega $ and reach their maximum as $\omega \rightarrow \omega _{\max }$.

We see that the asymptotic behavior of angular and polarization
distributions from one-electron (hole) state and from the vacuum state are
very similar. Nevertheless, one can distinguish between these two types of
the radiation. Indeed, probabilities (\ref{e32b}) and (\ref{e35}) are
proportional to macroscopic duration time $T,$ which is a consequence of the
integration over the large range $D$ of the momentum $p_{x}$ variation,
while probability (\ref{e29}) does not depend on $T$. Therefore, by studying
the dependence of the radiation on $T$, it is possible, in principle, to
identify its origin. In addition, one can stress that the main contribution
to the emission is due to probabilities (\ref{e32b}) and (\ref{e35}), if the
density of initial electrons (holes) is much less than density (\ref{n-cr})
of created electron-hole pairs.

The angular distribution is determined by the factor $\left\vert \chi
_{\vartheta }^{0,1}\right\vert ^{2}$ given by Eq. (\ref{e15}). It is quite
different for the polarization $\vartheta =1$ (polarization in the $XY$
plane) and for the polarization $\vartheta =2$ (polarization in the
perpendicular to vector $\boldsymbol{\epsilon }_{\mathbf{K}1}$ direction).
These factors are:
\begin{equation}
\left\vert \chi _{1}^{0,1}\right\vert ^{2}=\cos ^{2}\phi \ =\frac{k_{x}^{2}}{%
k_{x}^{2}+k_{y}^{2}},\;\left\vert \chi _{2}^{0,1}\right\vert ^{2}=\left\vert
\cos \theta \sin \phi \right\vert ^{2}=\frac{k_{z}^{2}k_{y}^{2}}{K^{2}\left(
k_{x}^{2}+k_{y}^{2}\right) }\ .  \label{e27b}
\end{equation}

Thus, the emission with the polarization $\vartheta =1$ has the same
probability along all the directions belonging to the plane which is
perpendicular to the one $XY${\large \ }and is tilted at the angle $\phi $\
with respect to the axis $x$. The maximum of the probability takes place for
a small angle $\phi $, $\cos ^{2}\phi \ \rightarrow 1$. The emission with
the polarization $\vartheta =1$ is absent in the $YZ$ plane, $\cos \phi =0$.
The emission with the polarization $\vartheta =2$ is also absent in the $XZ$
and $XY$ planes. The maximum of the emission is observed in the $YZ$ plane
in the directions close to $z$-axis, $\left\vert \cos \theta \sin \phi
\right\vert ^{2}\rightarrow 1$. Therefore, the emission in the $YZ$, $XZ$,
and $XY$ planes is highly polarized. We see that the unpolarized emission is
maximal in the directions defined by the relations $\cos ^{2}\phi \
\rightarrow 1$ and $\cos ^{2}\theta \rightarrow 1$. An emission in the $y$%
-axis, $\cos \phi =0$ and $\cos \theta =0$, is absent.

The above calculations of the emission are made in the first order of the
perturbation theory. This approximation is reasonable in case if total
emission probabilities from a given initial state are small. In this
relation, let us consider probability (\ref{e28}). One can see that
\begin{equation*}
\max \left. \left\vert \tilde{M}_{\mathbf{p}^{\prime }\mathbf{p}}^{\pm
}\right\vert ^{2}\right\vert _{\mathbf{p}^{\prime }=\mathbf{p}-\hbar \mathbf{%
k}}\sim 1,
\end{equation*}%
Integrating probability (\ref{e28}) all the frequencies in the domain where
high frequency approximation holds true, that is, from $\omega _{\min }$ to $%
\omega _{\mathrm{\max }}$ ($\omega _{\mathrm{\max }}$ is given by Eq. (\ref%
{e33})), we can estimate the maximum total emission probability from a
one-particle state. The smallness of this probability implies applicability
condition of the perturbation theory:%
\begin{equation}
\frac{\alpha }{\varepsilon }\left( \frac{v_{F}T}{c\Delta t_{st}}\right)
^{2}\ll 1.  \label{e38}
\end{equation}%
Since $v_{F}/c\sim 1/300$, condition (\ref{e38}) represents weak enough
restriction on the field parameter $T/\Delta t_{st}$\textrm{\ }satisfying
conditions~(\ref{time-condition}) and (\ref{g8}).

Let us consider probability (\ref{e35}). The function $\mathcal{R}\left(
\mathbf{K}\right) $ is restricted from above by the quantity $d\left( \omega
,0\right) $. Integrating probability (\ref{e35}) over{\large \ }the angles $%
\theta $\ and $\phi ${\large \ }and over the all frequencies in the domain
where high frequency approximation holds true, that is, from $\omega _{\min
} $ to $\omega _{\mathrm{\max }}$ ($\omega _{\min }$\ and $\omega _{\mathrm{%
\max }}$\ are given by Eqs. (\ref{e34}) and (\ref{e33}), respectively), we
can estimate the maximum total emission probability $\mathcal{P}_{\max }$
which accompanies the pair production from the vacuum state as:%
\begin{equation}
\mathcal{P}_{\max }\approx \frac{\alpha }{\varepsilon }\frac{2^{7/2}N_{f}S}{%
3l_{sc}^{2}}\left( \frac{T}{\Delta t_{st}}\right) ^{3}.  \label{e39a}
\end{equation}%
Probabilities (\ref{e35}) and (\ref{e39a}) grow with the increase of the
intensity of the electric field. At the same time, the total probability $%
P_{\max }$ grows especially fast due to the linear growth of the frequency
range. So, if the electric field increases by $q$ times, then the
probability $P_{\max }$ increases by $q^{5/2}$ times.

The smallness of probability{\large \ }(\ref{e39a}) implies also the
applicability condition of the perturbation theory:%
\begin{equation}
\mathcal{P}_{\max }\ll 1.  \label{e39b}
\end{equation}%
The typical quantity is $S\sim \left( 10^{-6}\mathrm{m}\right) ^{2}$. Using
estimation of $l_{sc}$ given by Eq. (\ref{l-scale}), we obtain $%
Sl_{sc}^{-2}\sim a\left( 48\right) ^{-2}$. This parameter can be considered
as a small one. In existing experiment conditions, where{\large \ }$T/\Delta
t_{st}${\large \ }\ and $a$\ \ satisfy conditions (\ref{time-condition}) and
(\ref{g9}), respectively,{\large \ }restriction (\ref{e39b}) may impose an
essential limit on the applicability of the perturbation theory.{\large \ }%
However, let us note that relation (\ref{g9}){\large \ }follows from the
assumption that during the time $T$ the electric field remains constant.
Assuming that the electric field $E$\ decreases quickly enough beyond the
formation interval, the upper limitation (\ref{g9}) to the intensity of the
electric field can be significantly weakened.\

\section{Summary\label{S5}}

In the present work, we have constructed an appropriate calculation
techniques in the framework of the reduced QED$_{3,2}$\ to describe one
species of Dirac fermions interacted with an external electric field and
photons in the graphene.\ In these techniques, effects of the vacuum
instability due to the particle creation by the external electric field are
taken into account nonperturbatively.\ In such a way, we consider the photon
emission in the graphene in the first-order approximation, taking into
account a vacuum instability by using the unitarity relation, and construct
closed formulas for total probabilities. Namely, we \ have calculated the
probabilities for the photon emission by an electron and for the photon
emission accompanying the vacuum instability in a quasiconstant electric
field that acts in the graphene plane during macroscopic time interval $T$.
In order to find corresponding mean values in the real graphene, results
obtained for one species of the Dirac fermions are multiplied by the number
of species $N_{f}=4$. It has been shown that the frequency of emission grows
in proportion to the duration of the electric field and reaches a final
maximum value before the electric field is turned off. The lower limit of
applicability of the perturbation theory with respect of the photon emission
is established and showed that the contribution of soft photons beyond this
boundary can be neglected.\ The obtained emission characteristics are
analyzed in a high frequency approximation which is more suitable for
possible experimental observations.\ The angular and polarization
distributions of the emission are also studied. The asymptotic behavior of
the unpolarized photon emission accompanying the vacuum instability matches
with the previous calculation in Ref. \cite{Yok14}, based on many-body
quantum mechanics.\ We see that the asymptotic behavior of angular and
polarization distributions from one-electron (hole) state and from the
vacuum state are very similar. Nevertheless, we point out that one can
distinguish between these two types of the radiation by considering the
emission under electric fields with different duration times $T$.\ The
applicability of the presented calculations to the graphene physics in
existing experimental conditions is shown.\ This implies also a general
possibility of laboratory verifying strong-field QED predictions, and, in
particular, real studying the Schwinger effect.

It was shown that in a high frequency approximation the\ variation\ of\ the\
external electric\ field\ acting\ on\ the\ particle\ within\ the\ formation\
length\ can\ be\ neglected,\ which justify the\ applicability of the
locally\ constant\ field\ approximation.\ Thus, the developed approach can
be easily extended to study the emission in any slowly varying field
configuration.

In the single graphene sheet there are actually two species of fermions in
the Dirac model of graphene. In our consideration, it is assumed that the
two cones of graphene are decupled and the system behaves like two copies of
a single Dirac cone. Topological insulators are characterized by a single
Dirac cone on each surface; see \cite{dassarma,top-insul11,top-insul11b} for
a review. Thus, the results obtained in the present study could be relevant
for a single Dirac cone on a surface of a topological insulator.

\bmhead{Acknowledgments}

The work is supported by Russian Science Foundation (Grant no. 19-12-00042).
Gitman is grateful to CNPq for continued support.

\section*{Declarations}

\begin{itemize}
\item Funding Not applicable

\item Conflict of interest/Competing interests (check journal-specific
guidelines for which heading to use) Not applicable

\item Ethics approval Not applicable

\item Consent to participate Not applicable

\item Consent for publication Not applicable

\item Data Availability Statement: No Data associated in the manuscript.

\item Code availability Not applicable

\item Authors' contributions Not applicable
\end{itemize}

\begin{appendices}

\section{Low frequency approximation \label{App2}}

Let us consider the probability densities, given by Eqs. (\ref{e9a}), (\ref%
{e9b}), and (\ref{e11}) in the range of low frequencies,%
\begin{equation}
u_{0}\ll 1.  \label{a21}
\end{equation}%
The ratio $\left\vert u_{x}\right\vert /u_{0}$\ , given by Eq. (\ref{lim5}),
is very small such that%
\begin{equation}
\rho \approx u_{0}\ll 1.  \label{a22}
\end{equation}%
In this limit, the behavior of the function $I_{j^{\prime },j}(\rho )$,
given by Eq. (\ref{e21c}), can be found using properties of the confluent
hypergeometric function $\Psi $; see Eqs. (6.8(2)) - (6.8(4)) from Ref. \cite%
{HTF1}. The only functions $I_{0,1}(\rho )$ and $I_{1,0}(\rho )$ grow as $%
\rho \rightarrow 0,$%
\begin{equation}
\left\vert I_{0,1}(\rho )\right\vert \sim \left\vert I_{1,0}(\rho
)\right\vert \sim \rho ^{-1}.  \label{a23}
\end{equation}%
Thus, the leading contribution to the amplitude $M_{\mathbf{p}^{\prime }%
\mathbf{p}}^{+}$ given by Eq. (\ref{e11}) is due to the terms $Y_{00}\approx
\mathcal{J}_{0,0}(\rho )\sim \rho ^{-1}$ and $Y_{11}\approx \mathcal{J}%
_{1,1}(\rho )\sim \rho ^{-1}$. The leading contribution to the amplitude $M_{%
\mathbf{p}^{\prime }\mathbf{p}}^{0}$ given by Eq. (\ref{e11}) is due to the
terms $\tilde{Y}_{01}\approx \mathcal{\tilde{J}}_{0,1}(\rho )\sim \rho ^{-1}$
and $\tilde{Y}_{10}\approx \mathcal{\tilde{J}}_{1,0}(\rho )\sim \rho ^{-1}$.
Therefore, the both modules squares amplitudes square grow proportionally to
$u_{0}^{-2}$,
\begin{equation}
\left. \left\vert M_{\mathbf{p}^{\prime }\mathbf{p}}^{0}\right\vert
^{2}\right\vert _{\mathbf{p}^{\prime }=-\mathbf{p}-\hbar \mathbf{k}}\sim
u_{0}^{-2},\;\left. \left\vert M_{\mathbf{p}^{\prime }\mathbf{p}}^{\pm
}\right\vert ^{2}\right\vert _{\mathbf{p}^{\prime }=\mathbf{p}-\hbar \mathbf{%
k}}\sim u_{0}^{-2}.  \label{a24}
\end{equation}%
At the same time, the both probability densities (\ref{e9a}) and (\ref{e9b})
are divergent functions of the order $u_{0}^{-1}$ as $u_{0}\rightarrow 0$,%
\begin{equation}
\frac{d\mathcal{P}\left( \mathbf{p};\left. \mathbf{K,}\vartheta \right\vert
0\right) }{du_{0}d\Omega }\sim \frac{\alpha }{\varepsilon }\left( \frac{v_{F}%
}{c}\right) ^{2}\frac{1}{u_{0}},\;\frac{d\mathcal{P}\left( \left. \mathbf{K}%
,\vartheta \right\vert \overset{\pm }{\mathbf{p}}\right) }{du_{0}d\Omega }%
\sim \frac{\alpha }{\varepsilon }\left( \frac{v_{F}}{c}\right) ^{2}\frac{1}{%
u_{0}}.  \label{a25}
\end{equation}

Such a behavior is an indication that the perturbative description of such
soft photons, does not work. Namely for photons with frequencies less than a
threshold frequency $u_{0}^{\mathrm{soft}}$, $u_{0}\lesssim u_{0}^{\mathrm{%
soft}}$, in case when functions (\ref{a25}) becomes of the order of unity.
This makes it possible to evaluate the quantity $u_{0}^{\mathrm{soft}}$,%
\begin{equation}
u_{0}^{\mathrm{soft}}\sim \frac{\alpha }{\varepsilon }\left( \frac{v_{F}}{c}%
\right) ^{2}.  \label{a26}
\end{equation}%
The number of such soft photons may be big enough. However, the only
physically measurable quantity is the emitted energy. This energy is a
negligibly small in the domain $u_{0}\lesssim u_{0}^{\mathrm{soft}}$. This
case is called the infrared catastrophe whose nature is associated with the
impossibility of separating a charged particle from its radiation field;
see, e.g., section 98 in Ref. \cite{BLP82} and sections 46 and 50.3 in Ref.
\cite{BogSh80}. The case of the strong-electric field QED is considered in
Ref. \cite{BFSh85}. The infrared divergences of QED are essentially
classical, and depend on the nature of the external current and on the
experimental resolution. The infrared catastrophe is absent from the
complete nonperturbative solution. Thus, one sees that the domain of the
applicability of the perturbation theory is $u_{0}>u_{0}^{\mathrm{soft}}$
and a contribution from the domain $u_{0}\lesssim u_{0}^{\mathrm{soft}}$ is
negligible.

In the case under consideration, the quantity $u_{0}^{\mathrm{soft}}$ is
very small, $u_{0}^{\mathrm{soft}}\sim 10^{-7}$. It follows from estimation (%
\ref{e33d}) and from restrictions on the parameter $u_{0}^{\mathrm{IR}}$
given by Eqs. (\ref{IR1}) and (\ref{IR2}) that in the realistic values of
the parameters $u_{0}^{\mathrm{soft}}\ll u_{0}^{\mathrm{IR}}$.

\section{Fourier transformation of the product of two WPC functions\label%
{App1}}

Integrals (\ref{e16}) can be reduced to a more simple form using the
Nikishov's representations given by Eq. (\ref{e17})) for the hyperbolic
coordinates $\rho $ and $\varphi $, see Ref. \cite{nikishov,nikishov79}. To demonstrate
how it works, we note that the integrals represent particular cases of the
more general integrals%
\begin{equation}
J_{\Lambda ^{\prime }\Lambda }^{\zeta ^{\prime }\zeta }\left( \rho ,\varphi
\right) =\int_{-\infty }^{+\infty }du\,f_{\Lambda ^{\prime }}^{\zeta
^{\prime }}(u-u_{x}/2)f_{\Lambda }^{\zeta }(u+u_{x}/2)e^{iu_{0}u}\,,
\label{a1}
\end{equation}%
where $f_{\Lambda }^{\zeta }(z)$ are WPCF's satisfying the differential
equation
\begin{equation}
\left( \frac{d^{2}}{dz^{2}}+z^{2}+\Lambda \right) f_{\Lambda }^{\zeta
}(z)=0\,,  \label{WPCFeq}
\end{equation}%
and $u_{0}$ and $u_{x}$, defined by Eq. (\ref{e10}), are:%
\begin{equation}
u_{0}=\rho \cosh \varphi ,\;u_{x}=\rho \sinh \varphi \;\mathrm{if}%
\;u_{0}^{2}>u_{x}^{2}.  \label{a2}
\end{equation}%
The functions $f_{\Lambda }^{\zeta }(z)$ with different values of $\zeta
=\pm $ are solutions of equation (\ref{WPCFeq}) with some complex parameters
$\Lambda $. In particular,%
\begin{eqnarray}
J_{\Lambda ^{\prime }\Lambda }^{-+}\left( \rho ,\varphi \right)
&=&Y_{j^{\prime }j},\;\Lambda =\lambda +i\left( 2j-1\right) ,\;\Lambda
^{\prime }=\lambda ^{\prime }+i\left( 1-2j^{\prime }\right) ,  \notag \\
J_{\Lambda ^{\prime }\Lambda }^{--}\left( \rho ,\varphi \right) &=&\tilde{Y}%
_{j^{\prime }j},\;\Lambda =\lambda +i\left( 1-2j\right) ,\;\Lambda ^{\prime
}=\lambda ^{\prime }+i\left( 1-2j^{\prime }\right) .  \label{a3}
\end{eqnarray}

Calculating the derivative of integral (\ref{a1}) with respect to the
hyperbolic angle $\varphi $, we find:%
\begin{eqnarray*}
&&\frac{\partial J_{\Lambda ^{\prime }\Lambda }^{\zeta ^{\prime }\zeta
}\left( \rho ,\varphi \right) }{\partial \varphi }=W+\int_{-\infty
}^{+\infty }iu_{x}f_{\Lambda ^{\prime }}^{\zeta ^{\prime
}}(u-u_{x}/2)f_{\Lambda }^{\zeta }(u+u_{x}/2)e^{iu_{0}u}du\ , \\
&&W=\frac{u_{0}}{2}\int_{-\infty }^{+\infty }\left[ f_{\Lambda ^{\prime
}}^{\zeta ^{\prime }}(u-u_{x}/2)\left. \frac{\partial \,f_{\Lambda }^{\zeta
}\left( z\right) }{\partial z}\right\vert _{z=u+u_{x}/2}\right. \\
&&\left. -\,\left. \frac{\partial \,f_{\Lambda ^{\prime }}^{\zeta ^{\prime
}}\left( z\right) }{\partial z}\right\vert _{z=u-u_{x}/2}f_{\Lambda }^{\zeta
}(u+u_{x}/2)\right] e^{iu_{0}u}du\ ,\ u_{x}=\frac{\partial u_{0}}{\partial
\varphi }\ ,\ u_{0}=\frac{\partial u_{x}}{\partial \varphi }\ .
\end{eqnarray*}%
Integrating by parts and neglecting boundary terms, we can transform $W$ to
the following form:%
\begin{eqnarray}
W &=&\frac{i}{2}\int_{-\infty }^{+\infty }\left[ f_{\Lambda ^{\prime
}}^{\zeta ^{\prime }}(u-u_{x}/2)\left. \frac{\partial \,^{2}f_{\Lambda
}^{\zeta }\left( z\right) }{\partial z^{2}}\right\vert _{z=u+u_{x}/2}\right.
\notag \\
&&\left. -\,\left. \frac{\partial ^{2}\,f_{\Lambda ^{\prime }}^{\zeta
^{\prime }}\left( z\right) }{\partial z^{2}}\right\vert
_{z=u-u_{x}/2}f_{\Lambda }^{\zeta }(u+u_{x}/2)\right] e^{iu_{0}u}du\ .
\label{a4}
\end{eqnarray}

Using equation (\ref{WPCFeq}) in integral (\ref{a4}), we find:%
\begin{equation}
\frac{\partial J_{\Lambda ^{\prime }\Lambda }^{\zeta ^{\prime }\zeta }\left(
\rho ,\varphi \right) }{\partial \varphi }=\frac{i}{2}\left( \Lambda
^{\prime }-\Lambda \right) J_{\Lambda ^{\prime }\Lambda }^{\zeta ^{\prime
}\zeta }\left( \rho ,\varphi \right) .  \label{a5}
\end{equation}%
Solutions of this equation are:%
\begin{equation}
J_{\Lambda ^{\prime }\Lambda }^{\zeta ^{\prime }\zeta }\left( \rho ,\varphi
\right) =e^{\frac{i}{2}\left( \Lambda ^{\prime }-\Lambda \right) \varphi
}J_{\Lambda ^{\prime }\Lambda }^{\zeta ^{\prime }\zeta }\left( \rho
,0\right) .  \label{a6}
\end{equation}%
We use the notation $J_{\Lambda ^{\prime }\Lambda }^{\zeta ^{\prime }\zeta
}\left( \rho \right) =J_{\Lambda ^{\prime }\Lambda }^{\zeta ^{\prime }\zeta
}\left( \rho ,0\right) $ in what follows. This function satisfies the
differential equation \cite{nikishov}%
\begin{equation}
\left[ \frac{d^{2}}{d\rho ^{2}}+\frac{1}{\rho }\frac{d}{d\rho }+\frac{\left(
\Lambda -\Lambda ^{\prime }\right) ^{2}}{4\rho ^{2}}+\frac{\rho ^{2}}{4}-%
\frac{\Lambda +\Lambda ^{\prime }}{2}\right] J_{\Lambda ^{\prime }\Lambda
}^{\zeta ^{\prime }\zeta }\left( \rho \right) =0\,.  \label{a7}
\end{equation}%
This fact can be verified performing integrations by parts with account
taken of Eq. (\ref{WPCFeq}). The differential equation (\ref{a7}) can be
reduced to a confluent hypergeometric equation. Using two linearly
independent solutions of such an equation, we find general solution of the
differential equation (\ref{a7})%
\begin{eqnarray}
&&J_{\Lambda ^{\prime }\Lambda }^{\zeta ^{\prime }\zeta }\left( \rho \right)
=e^{-\eta /2}\left[ C_{1}\eta ^{i\beta }\Phi \left( \frac{i\Lambda }{2}+%
\frac{1}{2},1+2i\beta ;\eta \right) +C_{2}\eta ^{-i\beta }\Phi \left( \frac{%
i\Lambda ^{\prime }}{2}+\frac{1}{2},1-2i\beta ;\eta \right) \right] ,  \notag
\\
&&\eta =-i\rho ^{2}/2,\;\beta =\left( \Lambda -\Lambda ^{\prime }\right)
/4\;,  \label{a8}
\end{eqnarray}%
where the $C_{1}$ and $C_{2}$ are some undetermined coefficients, which must
be fixed by appropriate boundary conditions, so that solution (\ref{a8})
corresponds to the original integral (\ref{a1}).

The confluent hypergeometric function $\Phi \left( a,c;\eta \right) $ is
entire in $\eta $ and $a$, and is a meromorphic function of $c$. Note that $%
\Phi \left( a,c;0\right) =1$. WPCF's are entire functions of $\Lambda $ and $%
\Lambda ^{\prime }$. One can see that the integrals $J_{\Lambda ^{\prime
}\Lambda }^{\zeta ^{\prime }\zeta }\left( \rho \right) $ are entire
functions of $\Lambda $ and $\Lambda ^{\prime }$ and meromorphic functions
of $\Lambda -\Lambda ^{\prime }$. Then one can find a boundary condition $%
J_{\Lambda ^{\prime }\Lambda }^{\zeta ^{\prime }\zeta }\left( \rho \right) $
at $\rho \rightarrow 0$ for some convenient values of $j$ and $j^{\prime }$.
The remaining integrals $J_{\Lambda ^{\prime }\Lambda }^{\zeta ^{\prime
}\zeta }\left( \rho \right) $ can be obtained extending domains of $\Lambda $
and $\Lambda ^{\prime }$ by an analytic continuation.

Let us start with\emph{\ }$\tilde{J}_{0,0}(\rho )$\emph{\ }given by Eq. (\ref%
{e18b}). This integral can be represented as a solution of equation (\ref{a7}%
) where $\Lambda ^{\prime }=\lambda ^{\prime }+i$ and $\Lambda =\lambda +i$.
The coefficients $C_{1}$ and $C_{2}$ in Eq. (\ref{a8}) can be fixed by a
comparison with the $\rho \rightarrow 0$ limit of integral~(\ref{e18b}). Let
us first represent this integral as follows:%
\begin{eqnarray}
&&\mathcal{\tilde{J}}_{0,0}(\rho
)=F^{0}+F^{+}+F^{-},\;F^{+}=\int_{0}^{\infty }f^{+}\left( u\right) e^{i\rho
u}du,\;F^{-}=\int_{-\infty }^{0}f^{-}\left( u\right) e^{i\rho u}du\ ,  \notag
\\
&&F^{0}=\int_{0}^{\infty }f\left( u\right) \left[ f\left( u\right)
-f^{+}\left( u\right) \right] e^{i\rho u}du+\int_{-\infty }^{0}f\left(
u\right) \left[ f\left( u\right) -f^{-}\left( u\right) \right] e^{i\rho
u}du\ ,  \notag \\
&&f\left( u\right) =D_{-\nu ^{\prime }}[-(1+i)u]D_{-\nu }[-(1+i)u]\ ,
\label{a9}
\end{eqnarray}%
where $f^{\pm }\left( u\right) =\left. f\left( u\right) \right\vert
_{u\rightarrow \pm \infty }$. It can be seen that function (\ref{a8}) is
reduced to the oscillations $C_{1}\eta ^{i\beta }+C_{2}\eta ^{-i\beta }$ as $%
\rho \rightarrow 0$. Then $\rho $-independent terms do not contribute to the
integrals $F^{0}$ and $F^{\pm }$. Taking into account that $\lim_{\rho
\rightarrow 0}F^{0}$ and $\lim_{\rho \rightarrow 0}F^{-}$ do not depend on $%
\rho ,$ one sees that the oscillation terms of $F^{+}$ are only essential.
Using relations (8.2.(7)) and (8.4.(1)) from Ref. \cite{HTF2}, one finds:%
\begin{eqnarray}
\mathcal{\tilde{J}}_{0,0}(\rho ) &=&\sqrt{\pi }e^{i\pi \left( \nu +\nu
^{\prime }-1\right) /4}\left[ e^{i\pi \nu /2}\frac{\Gamma \left( \nu -\nu
^{\prime }\right) }{\Gamma \left( \nu \right) }\left( \frac{\rho }{\sqrt{2}}%
\right) ^{\nu ^{\prime }-\nu }\right.  \notag \\
&&\left. +e^{i\pi \nu ^{\prime }/2}\frac{\Gamma \left( \nu ^{\prime }-\nu
\right) }{\Gamma \left( \nu ^{\prime }\right) }\left( \frac{\rho }{\sqrt{2}}%
\right) ^{\nu -\nu ^{\prime }}\;\mathrm{as}\;\rho \rightarrow 0\right] .
\label{a10}
\end{eqnarray}%
Comparing Eqs. (\ref{a8}) and (\ref{a10}), we obtain:%
\begin{equation}
C_{1}=\sqrt{\pi }e^{i\pi \left( \nu +\nu ^{\prime }-1/2\right) /2}\frac{%
\Gamma \left( \nu ^{\prime }-\nu \right) }{\Gamma \left( \nu ^{\prime
}\right) },\;C_{2}=\sqrt{\pi }e^{i\pi \left( \nu +\nu ^{\prime }-1/2\right)
/2}\frac{\Gamma \left( \nu -\nu ^{\prime }\right) }{\Gamma \left( \nu
\right) }.  \label{a11}
\end{equation}%
Using relation (6.5.(7)) from Ref. \cite{HTF1}, one can represent function
given by Eqs. (\ref{a8}) and (\ref{a11}) as%
\begin{equation}
\mathcal{\tilde{J}}_{0,0}(\rho )=\sqrt{\pi }e^{i\pi \left( \nu +\nu ^{\prime
}-1/2\right) /2}e^{-\eta /2}\eta ^{\left( \nu -\nu ^{\prime }\right) /2}\Psi
\left( \nu ,1+\nu -\nu ^{\prime };\eta \right) ,  \label{a12}
\end{equation}%
where $\Psi \left( \nu ,1+\nu -\nu ^{\prime };\eta \right) $ is the
confluent hypergeometric function,%
\begin{eqnarray}
\Psi \left( \nu ,1+\nu -\nu ^{\prime };\eta \right) &=&\frac{\Gamma \left(
\nu ^{\prime }-\nu \right) }{\Gamma \left( \nu ^{\prime }\right) }\Phi
\left( \nu ,1+\nu -\nu ^{\prime };\eta \right)  \notag \\
&&+\frac{\Gamma \left( \nu -\nu ^{\prime }\right) }{\Gamma \left( \nu
\right) }\eta ^{\nu ^{\prime }-\nu }\Phi \left( \nu ^{\prime },1+\nu
^{\prime }-\nu ;\eta \right) .  \label{a13}
\end{eqnarray}%
Using transformation $\nu \rightarrow \nu +j$ and $\nu ^{\prime }\rightarrow
\nu ^{\prime }+j^{\prime }$ in Eq. (\ref{a12}), one obtains the final form (%
\ref{e21a}) for integral (\ref{e18b}).

The integral $\mathcal{J}_{j^{\prime },j}(\rho )$ given by Eq. (\ref{e18a})
can be represented as the solution of equation (\ref{a7}) where $\Lambda
^{\prime }=\lambda ^{\prime }+i\left( 1-2j^{\prime }\right) $ and $\Lambda
=\lambda +i\left( 2j-1\right) $. Using relation (8.2.(6)) from Ref. \cite%
{HTF2}, we transform one of the WPCF's in Eq. (\ref{e18a}) to obtain
convenient representations:
\begin{eqnarray}
&&\mathcal{J}_{j^{\prime },j}(\rho )=\frac{\Gamma \left( \nu -j+1\right) }{%
\sqrt{2\pi }}\left[ e^{i\pi \left( \nu -j\right) /2}\mathcal{\tilde{J}}%
_{j^{\prime },1-j}(\rho )+e^{-i\pi \left( \nu -j\right) /2}\mathcal{J}%
_{j^{\prime },1-j}^{\prime }(\rho )\right] \ ,  \label{a15a} \\
&&\mathcal{J}_{j^{\prime },1-j}^{\prime }(\rho )=\int_{-\infty }^{\infty
}D_{-\nu ^{\prime }-j^{\prime }}[-(1+i)u]D_{-\nu +j-1}[(1+i)u]e^{i\rho u}du\
,  \label{a15b}
\end{eqnarray}%
where $\mathcal{\tilde{J}}_{j^{\prime },1-j}(\rho )$ is given by Eq. (\ref%
{e21a}). The integral $\mathcal{J}_{j^{\prime },1-j}^{\prime }(\rho )$ is
represented by function (\ref{a8}) where some coefficients $C_{1}^{\prime }$
and $C_{2}^{\prime }$ can be fixed by the comparison with $\rho \rightarrow
0 $ limit of the integral $\mathcal{J}_{j^{\prime },1-j}^{\prime }(\rho )$.

Let us start with $\mathcal{J}_{0,0}^{\prime }(\rho )$, where $\Lambda
^{\prime }=\lambda ^{\prime }+i$ and $\Lambda =\lambda +i$. In this case, it
can be seen that function (\ref{a8}) takes the form $C_{1}^{\prime }\eta
^{i\beta }+C_{2}^{\prime }\eta ^{-i\beta }$ as $\rho \rightarrow 0$. Hence
all $\rho $-independent terms of $\mathcal{J}_{0,0}^{\prime }(\rho )$, given
by Eq. (\ref{a15b}), can be ignored at $\rho \rightarrow 0$ limit and only
the oscillation terms of following integrals
\begin{eqnarray}
&&G^{+}=\int_{0}^{\infty }g^{+}\left( u\right) e^{i\rho
u}du,\;F^{-}=\int_{-\infty }^{0}g^{-}\left( u\right) e^{i\rho u}du,\;g^{\pm
}\left( u\right) =\left. g\left( u\right) \right\vert _{u\rightarrow \pm
\infty }\ ,  \notag \\
&&g\left( u\right) =D_{-\nu ^{\prime }}[-(1+i)u]D_{-\nu }[(1+i)u]
\label{a16}
\end{eqnarray}%
are essential. Using relations (8.2.(7)) and (8.4.(1)) from Ref. \cite{HTF2}%
, one finds:
\begin{eqnarray}
\mathcal{J}_{0,0}^{\prime }(\rho ) &=&\sqrt{\pi }e^{i\pi \left( \nu ^{\prime
}-\nu -1/2\right) /2}\left[ e^{i\pi \left( \nu -\nu ^{\prime }\right) /4}%
\frac{\Gamma \left( \nu -\nu ^{\prime }\right) }{\Gamma \left( \nu \right) }%
\left( \frac{\rho }{\sqrt{2}}\right) ^{\nu ^{\prime }-\nu }\right.   \notag
\\
&&\left. +e^{-i\pi \left( \nu -\nu ^{\prime }\right) /4}\frac{\Gamma \left(
\nu ^{\prime }-\nu \right) }{\Gamma \left( \nu ^{\prime }\right) }\left(
\frac{\rho }{\sqrt{2}}\right) ^{\nu -\nu ^{\prime }}\;\mathrm{as}\;\rho
\rightarrow 0\right] .  \label{a17}
\end{eqnarray}%
Comparing Eqs. (\ref{a8}) and (\ref{a17}) we obtain:%
\begin{equation}
C_{1}^{\prime }=\sqrt{\pi }e^{i\pi \left( \nu ^{\prime }-\nu -1/2\right) /2}%
\frac{\Gamma \left( \nu ^{\prime }-\nu \right) }{\Gamma \left( \nu ^{\prime
}\right) },\;C_{2}^{\prime }=\sqrt{\pi }e^{i\pi \left( \nu ^{\prime }-\nu
-1/2\right) /2}\frac{\Gamma \left( \nu -\nu ^{\prime }\right) }{\Gamma
\left( \nu \right) }.  \label{a18}
\end{equation}%
Using relation (6.5.(7)) from Ref. \cite{HTF1}, the function given by Eqs. (%
\ref{a8}) and (\ref{a18}) can be represented as:%
\begin{equation}
\mathcal{J}_{0,0}^{\prime }(\rho )=\sqrt{\pi }e^{i\pi \left( \nu ^{\prime
}-\nu -1/2\right) /2}e^{-\eta /2}\eta ^{\left( \nu -\nu ^{\prime }\right)
/2}\Psi \left( \nu ,1+\nu -\nu ^{\prime };\eta \right) .  \label{a19}
\end{equation}%
Using the transformations $\nu \rightarrow \nu +1-j$ and $\nu ^{\prime
}\rightarrow \nu ^{\prime }+j^{\prime }$ in Eq. (\ref{a19}), we obtain the
following representation for integral (\ref{a15b}):%
\begin{equation}
\mathcal{J}_{j^{\prime },1-j}^{\prime }(\rho )=e^{-i\pi \left( \nu -\nu
^{\prime }+1-j-j^{\prime }\right) /2}I_{j^{\prime },1-j}(\rho ),  \label{a20}
\end{equation}%
where $I_{j^{\prime },j}(\rho )$ is given by Eq. (\ref{e21c}). Substituting
representations (\ref{e21a}) and (\ref{a20}) into Eq. (\ref{a15a}) we find
the final form (\ref{e21b}).

\end{appendices}


\bibliographystyle{plain}
\bibliography{sn-bibliography}


\end{document}